\documentclass{pasj01}
\usepackage{listings}
\usepackage{color}
\newcommand{\myvec}[1]{\vec{#1}}

\newcommand{\icarus}{Icarus}

\begin{document}

\title{Implementation and Performance of FDPS: A Framework for Developing
  Parallel Particle Simulation Codes}

\author{Masaki \textsc{Iwasawa}\altaffilmark{1}}
\email{masaki.iwasawa@riken.jp}
\altaffiltext{1}{RIKEN Advanced Institute for Computational Science,
7--1--26 Minatojima-minami-machi, Chuo-ku, Kobe, Hyogo, Japan}

\author{Ataru \textsc{Tanikawa}\altaffilmark{1,2}}
\email{tanikawa@ea.c.u-tokyo.ac.jp}
\altaffiltext{2}{Department of Earth and Astronomy, College of Arts and Science, The University of
Tokyo, 3--8--1 Komaba, Meguro-ku, Tokyo, Japan}
  
\author{Natsuki \textsc{Hosono}\altaffilmark{1}}
\email{natsuki.hosono@riken.jp}

\author{Keigo \textsc{Nitadori}\altaffilmark{1}}
\email{keigo@riken.jp}

\author{Takayuki \textsc{Muranushi}\altaffilmark{1}}
\email{takayuki.muranushi@riken.jp}

\author{Junichiro \textsc{Makino}\altaffilmark{3,1,4}}
\email{makino@mail.jmlab.jp}
\altaffiltext{3}{Department of Planetology, Graduate School of
  Science, Kobe University, 1-1, Rokkodai-cho, Nada-ku, Kobe, Hyogo,
  Japan}
\altaffiltext{4}{Earth-Life Science Institute, Tokyo Institute of
  Technology, 2--12--1 Ookayama, Meguro-ku, Tokyo, Japan}


\maketitle

\begin{abstract}

We present the basic idea, implementation, measured performance and
performance model of FDPS (Framework for developing particle
simulators). FDPS is an application-development framework which helps
the researchers to develop simulation programs using particle methods
for large-scale distributed-memory parallel supercomputers. A
particle-based simulation program for distributed-memory parallel
computers needs to perform domain decomposition, exchange of particles
which are not in the domain of each computing node, and gathering of
the particle information in other nodes which are necessary for
interaction calculation. Also, even if distributed-memory parallel
computers are not used, in order to reduce the amount of computation,
algorithms such as Barnes-Hut tree algorithm or Fast Multipole Method
should be used in the case of long-range interactions. For short-range
interactions, some methods to limit the calculation to neighbor
particles are necessary. FDPS provides all of these functions which
are necessary for efficient parallel execution of particle-based
simulations as ``templates'', which are independent of the actual data
structure of particles and the functional form of the
particle-particle interaction. By using FDPS, researchers can write
their programs with the amount of work necessary to write a simple,
sequential and unoptimized program of $O(N^2)$ calculation cost, and
yet the program, once compiled with FDPS, will run efficiently on
large-scale parallel supercomputers. A simple gravitational $N$-body
program can be written in around 120 lines. We report the actual
performance of these programs and the performance model. The weak
scaling performance is very good, and almost linear speedup was
obtained for up to the full system of K computer. The minimum
calculation time per timestep is in the range of 30 ms ($N=10^7$) to
300 ms ($N=10^9$). These are currently limited by the time for the
calculation of the domain decomposition and communication necessary
for the interaction calculation. We discuss how we can overcome these
bottlenecks.

\end{abstract}

\section{Introduction}

In the field of computational astronomy, simulations based on particle
methods have been widely used. In such simulations, a system is either
physically a collection of particles as in the case of star clusters,
galaxies and dark-matter halos, or modeled by a collection of
particles, as in SPH (smoothed particle hydrodynamics) simulation of
astrophysical fluids.  Since particles move automatically as the
result of integration of the equation of motion of the particle,
particle-based simulations have an advantage for systems experiencing
strong deformation or systems with high density contrast.  This is one
of the reasons why particle-based simulations are widely used in
astronomy. Examples of particle-based simulations include cosmological
simulations or planet-formation simulations with gravitational
$N$-body code, simulations of star and galaxy formation with the SPH
code or other particle-based codes, and simulations of planetesimal
formation with the DEM (discrete element method) code.

We need to use a large number of particles to improve the resolution
and accuracy of particle-based simulations, and in order to do so we
need to increase the calculation speed and need to use
distributed-memory parallel machines efficiently. In other words, we
need to implement efficient algorithms such as the Barnes-Hut tree
algorithm \citep{1986Natur.324..446B}, the TreePM
algorithm \citep{1995ApJS...98..355X} or the Fast Multipole Method
\citep{2000ApJ...536L..39D} to distributed-memory parallel computers.
In order to achieve high efficiency, we need to divide a computational
domain into subdomains in such a way that minimizes the need of
communication between processors to maintain the division and to
perform interaction calculations. To be more specific, parallel
implementations of particle-based simulations contain the following
three procedures to achieve the high efficiency: (a) domain
decomposition, in which the subdomains to be assigned to computing
nodes are determined so that the calculation times are balanced, (b)
particle exchange, in which particles are moved to computing nodes
corresponding to the subdomains to which they belong, and (c)
interaction information exchange, in which each computing node
collects the information necessary to calculate the interactions on
its particles.  In addition, we need to make use of multiple CPU cores
in one processor chip and SIMD (single instruction multiple data)
execution units in one CPU core, or in some cases GPGPUs
(general-purpose computing on graphics processing units) or other
accelerators.

In the case of gravitational $N$-body problems, there are a number of
works in which the efficient parallelization is
discussed \citep{1994JCoPh.111..136S, 1996NewA....1..133D,
2004PASJ...56..521M, 2009PASJ...61.1319I,
Ishiyama:2012:PAN:2388996.2389003}.  The use of SIMD units is
discussed in \citet{2006NewA...12..169N}, \citet{2012NewA...17...82T}
and \citet{2013NewA...19...74T}, and GPGPUs
in \citet{2009NewA...14..630G}, \citet{hamada2009novel}, \citet{Hamada:2009:THN:1654059.1654123}, \citet{Hamada:2010:TAN:1884643.1884644}, \citet{2012JCoPh.231.2825B}
and
\citet{Bedorf:2014:PGT:2683593.2683600}.

In the field of molecular dynamics, several groups have been working
on parallel implementations. Examples of such efforts include
Amber \citep{2015AMBER}, CHARMM \citep{2009CHARMM},
Desmond \citep{GB14}, GROMACS \citep{2014GROMACS},
LAMMPS \citep{1995LAMMPS}, NAMD \citep{2005NAMD}. In the field of CFD
(computational fluid dynamics), Many commercial and non-commercial
packages now support SPH or other particle-based methods
(PAM-CRASH \footnote{https://www.esi-group.com/pam-crash},
LS-DYNA \footnote{http://www.lstc.com/products/ls-dyna},
Adventure/LexADV \footnote{http://adventure.sys.t.u-tokyo.ac.jp/lexadv})

Currently, parallel application codes are being developed for each of
specific applications of particle methods. Each of these codes
requires multi-year effort of a multi-person team. We believe this
situation is problematic because of the following reasons.

First, it has become difficult for researchers to try a new method or
just a new experiment which requires even a small modification of
existing large codes. If one wants to test a new numerical scheme, the
first thing he or she would do is to write a small program and to do
simple tests. This can be easily done, as far as that program runs on
one processor. However, if he or she then wants to try a
production-level large calculation using the new method, the
parallelization for distributed-memory machines is necessary, and
other optimizations are also necessary. However, to develop such a
program in a reasonable time is impossible for a single person, or
even for a team, unless they already have experiences of developing
such a code.

Second, even for a team of people developing a parallel code for one
specific problem, it has become difficult to take care of all the
optimizations necessary to achieve a reasonable efficiency on recent
processors. In fact, the efficiency of many simulation codes mentioned
above on today's latest microprocessors are rather poor, simply
because the development team does not have enough time and expertise
to implement necessary optimizations (in some case they require the
change of data structure, control structure and algorithms).

In our opinion, these difficulties have significantly slowed down
researchs in the numerical methods and also the research using
large-scale simulations.

We have developed FDPS (Framework for Developing Particle
Simulator)\footnote{https://github.com/FDPS/FDPS} \citep{2015FDPS} in
order to solve these difficulties. The goal of FDPS is to let
researchers concentrate on the implementation of numerical schemes and
physics, without spending too much time on parallelization and code
optimization. To achieve this goal, we separate a code into domain
decomposition, particle exchange, interaction information exchange and
fast interaction calculation using Barnes-Hut tree algorithm and/or
neighbor search and the rest of the code. We implement these functions
as ``template'' libraries in C++ language. The reason why we use the
template libraries is to allow the researchers to define their own
data structure for particles and their own functions for
particle-particle interactions, and to provide them with
highly-optimized libraries with small software overhead.  A user of
FDPS needs to define the data structure and the function to evaluate
particle-particle interaction. Using them as template arguments, FDPS
effectively generates the highly-optimized library functions which
perform complex operations listed above.

From users' point of view, what is necessary is to write the program
in C++, using FDPS library functions and to compile it using a
standard C++ compiler. Using FDPS, users thus can write their
particle-based simulation codes for gravitational $N$-body problem,
SPH, MD (molecular dynamics), DEM, or many other particle-based
methods, without spending their time on parallelization and complex
optimization. The compiled code will run efficiently on large-scale
parallel machines.

For grid-based or FEM (Finite Element Method) applications, there are
many frameworks for developing parallel applications. For example,
Cactus \citep{2003Cactus} has been widely used for numerical
relativity, and BoxLib \footnote{https://ccse.lbl.gov/BoxLib/} is
designed for AMR (adaptive mesh refinement). For particle-based
simulations, such frameworks have not been widely used yet, though
there were early efforts as in \citet{1995CoPhC..87..266W}, which is
limited to long-range, $1/r$ potential. More recently, LexADV\_EMPS is
currently being developed \citep{2015LexADV_EMPS}. As its name
suggests, it is rather specialized to the EMPS (Explicit Moving
Particle Simulation) method \citep{2014Murotani}.

In section~\ref{sec:user}, we describe the basic design concept of
FDPS.  In section~\ref{sec:implementation}, we describe the
implementation of parallel algorithms in FDPS. In
section~\ref{sec:performance}, we present the measured performance for
three astrophysical applications developed using FDPS. In
section~\ref{sec:performancemodel}, we construct a performance model
and predict the performance of FDPS on near-future supercomputers.
Finally, we summarize this study in section~\ref{sec:conclusion}.

\section{How FDPS works}
\label{sec:user}

In this section, we describe the design concept of FDPS. In
section~\ref{sec:design}, we present the design concept of FDPS. In
section~\ref{sec:samplecode}, we show an $N$-body simulation code
written using FDPS, and describe how FDPS is used to perform
parallelization algorithms. Part of the contents in this scetion have been
published in \citet{2015FDPS}.

\subsection{Design concept}
\label{sec:design}

In this section, we present the basic design concept of FDPS. We first
present the abstract view of calculation codes for particle-based
simulations on distributed-memory parallel computers, and then
describe how such abstraction is realized in FDPS.

\subsubsection{Abstract view of particle-based simulation codes}
\label{sec:view}

In a particle-based simulation code that uses the space decomposition on
distributed-memory parallel computers, the calculation proceeds in the
following steps:
\begin{enumerate}

\item The computational domain is divided into subdomains, and each
  subdomain is assigned to one MPI process. This step is usually
  called domain decomposition. Here, minimization of inter-process
  communication and a good load balance among  processes should be
  achieved.
 \label{proc:decompose}

\item Particles are exchanged among  processes, so that each
  process owns particles in its subdomain. In this paper we call this
  step particle exchange.
  \label{proc:exchange}

\item Each process collects the information necessary to calculate the
  interactions on its particles. We call this step
  interaction-information exchange.
  \label{proc:interactionexchange}

\item Each process calculates interactions between particles in its
  subdomain. We call this step interaction calculation.
  \label{proc:interaction}

\item The data for each particle are updated using the calculated
  interactions. This part is done without inter-process communication.
   \label{proc:local}
\end{enumerate}

Steps \ref{proc:decompose}, \ref{proc:exchange}, and
\ref{proc:interactionexchange} involve parallelization and
inter-process communications. FDPS provides library functions to
perform these parts. Therefore, users of FDPS do not have to write the
parallelization and/or inter-process communication part of their own
code at all.

Step \ref{proc:interaction} does not involve inter-process
communication. However, this step are necessary to perform the actual
calculation of interactions between two particles.  Users of FDPS
should write a simple interaction kernel. The actual interaction
calculation using the tree algorithm or neighbor search is done in the
FDPS side.

Step \ref{proc:local} involves neither inter-process communication nor
interaction calculation. Users of FDPS should and can write their own
program for this part. 

FDPS can be used to implement particle-based simulation codes for
initial value problems which can be expressed as the following
ordinary differential equations:

\begin{eqnarray}
  \frac{d\myvec{u}_i}{dt} = \myvec{g}\left(\sum_j^N \myvec{f}
  (\myvec{u}_i, \myvec{u}_j), \myvec{u}_i\right). \label{eq:geq}
\end{eqnarray}
  
Here, $N$ is the number of particles in the system, $\myvec{u}_i$ is
a vector which represents the physical quantities of  particle $i$, 
$\myvec{f}$ is a function which describes
the contribution of particle $j$ to the time derivative of
physical quantities of particle $i$, and $\myvec{g}$ is a function
which converts the sum of the contributions to the actual time
derivative. In the case of gravitational $N$-body
simulation, $\myvec{u}_i$ contains position, velocity, mass, and other
parameters of  particle $i$, $\myvec{f}$ is the gravitational force
from particle $j$ to particle $i$, and 
$\myvec{g}$ gives velocity as the time derivative of  position and
calculated acceleration as the time derivative of velocity.

Hereafter, we call a particle that receives the interaction
``$i$-particle'', and a particle exerting that interaction
``$j$-particle''. The actual contents of vector $\myvec{u}_i$ and the
functional forms of $\myvec{f}$ and $\myvec{g}$ depend on the physical
system and numerical scheme used.

In equation (\ref{eq:geq}) we included only the pairwise interactions,
because usually the calculation cost of the pairwise interaction is
the highest even when many-body interaction is important. For example,
angle and torsion of bonding force in simulation of organic molecules
can be done in the user code, with small additional computing cost.

\subsubsection{Design concept of FDPS}
\label{sec:concept}

In this section, we describe how the abstract view presented in the
previous section is actually expressed in the FDPS API (application
programming interface).  The API of FDPS is defined as a set of
template library functions in C++ language.

\begin{figure}
  \begin{center}
    \includegraphics[width=8cm]{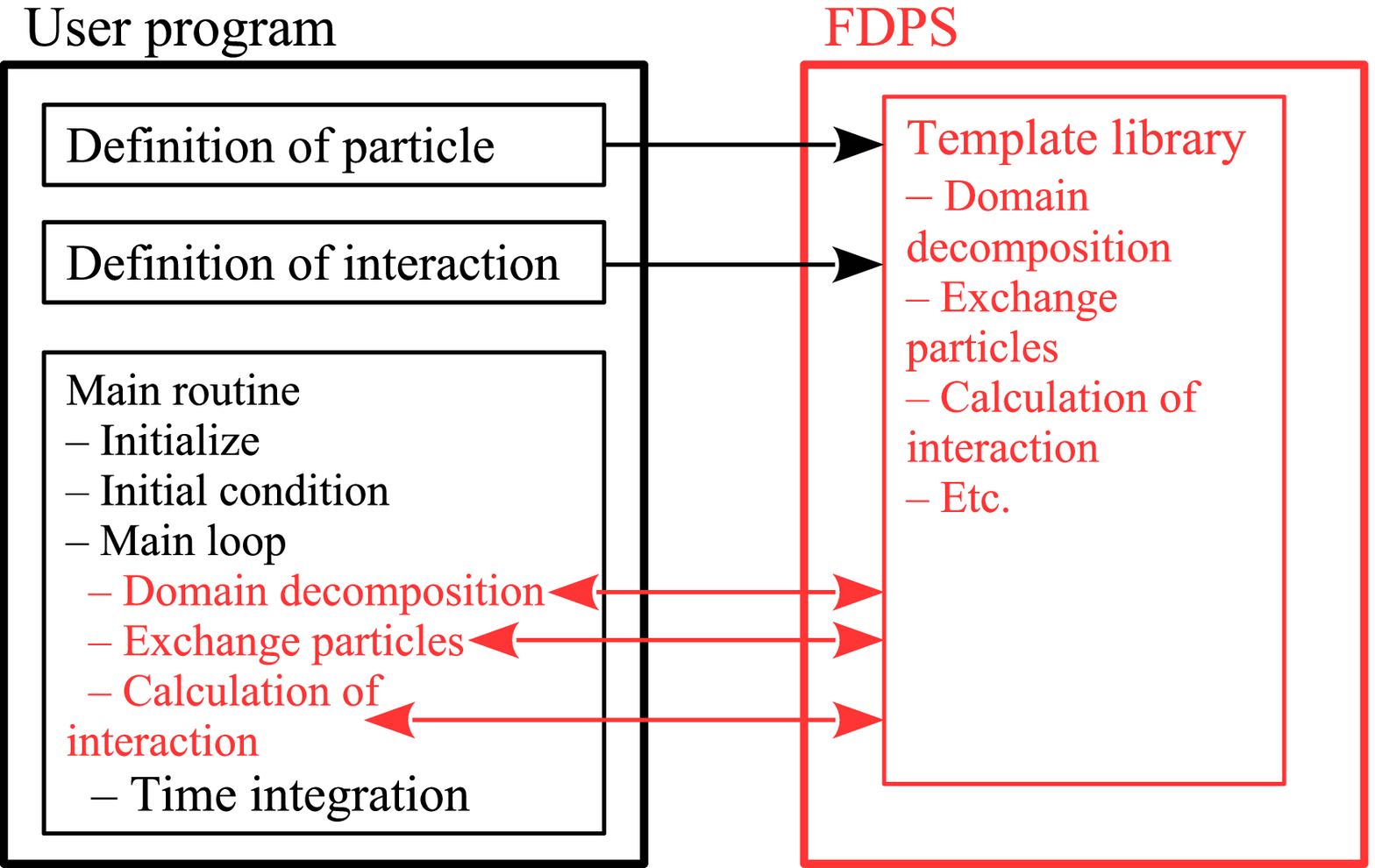}
  \end{center}
  \caption{The basic concept of FDPS. The user program gives the
    definitions of particle and interaction to FDPS, and calls FDPS
    APIs.}
  \label{fig:concept}
\end{figure}

Figure~\ref{fig:concept} shows how a user program and FDPS library
functions interact.  The user program gives the definition of a
particle $\myvec{u}_i$ and particle-particle interaction $\myvec{f}$
to FDPS at the compile time. They are written in the standard C++
language. Thus, the user program [at least the main() function]
currently should be written in C++\footnote{We will investigate a
  possible way to use APIs of FDPS from programs written in Fortran.}.

The user program first does the initialization of FDPS library. When
the user program is compiled for the MPI environment, the
initialization of MPI communication is done in the FDPS initialization
function.  The setup of the initial condition is done in the user
program.  It is possible to use file input function defined in FDPS.
In the main integration loop, domain decomposition, particle exchange,
interaction information exchange and force calculation are all done
through library calls to FDPS.  The time integration of the physical
quantities of particles using the calculated interaction, is done
within the user program.

Note that it is possible to implement multi-stage integration schemes
such as the Runge-Kutta schemes using FDPS. FDPS can evaluate the
right-hand side of equation (\ref{eq:geq}) for a given set of
$\myvec{u}_i$. Therefore, the derivative calculation for intermediate
steps can be done by passing $\myvec{u}_i$ containing appropriate
values.

The parallelization using MPI is completely taken care by FDPS, and
the use of OpenMP is also taken care by FDPS for the interaction
calculation. In order to achieve high performance, the interaction
calculation should make efficient use of the cache memory and SIMD
units. In FDPS, this is done by requiring an interaction calculation
function to calculate the interactions between multiple $i$- and
$j$-particles. In this way, the amount of memory access is
significantly reduced, since single $j$-particle is used to calculate
the interaction on multiple $i$-particles ($i$-particles are also in
the cache memory). To make the efficient use of the SIMD execution
units, currently the user should write the interaction calculation
loop so that the compiler can generate SIMD instructions. Of course,
the use of libraries optimized for specific architectures
\citep{2006NewA...12..169N, 2012NewA...17...82T, 2013NewA...19...74T}
would guarantee very high performance.

It is also possible to use GPUs and other accelerators for the
interaction calculation. In order to reduce the communication
overhead, so-called ``multiwalk'' method \citet{hamada2009novel}, is
implemented. Thus, interaction calculation kernels for accelerators
should take multiple sets of the pair of $i$- and $j$-particles. The
performance of this version will be discussed elsewhere.

\begin{figure}
  \begin{center}
    \includegraphics[width=8cm]{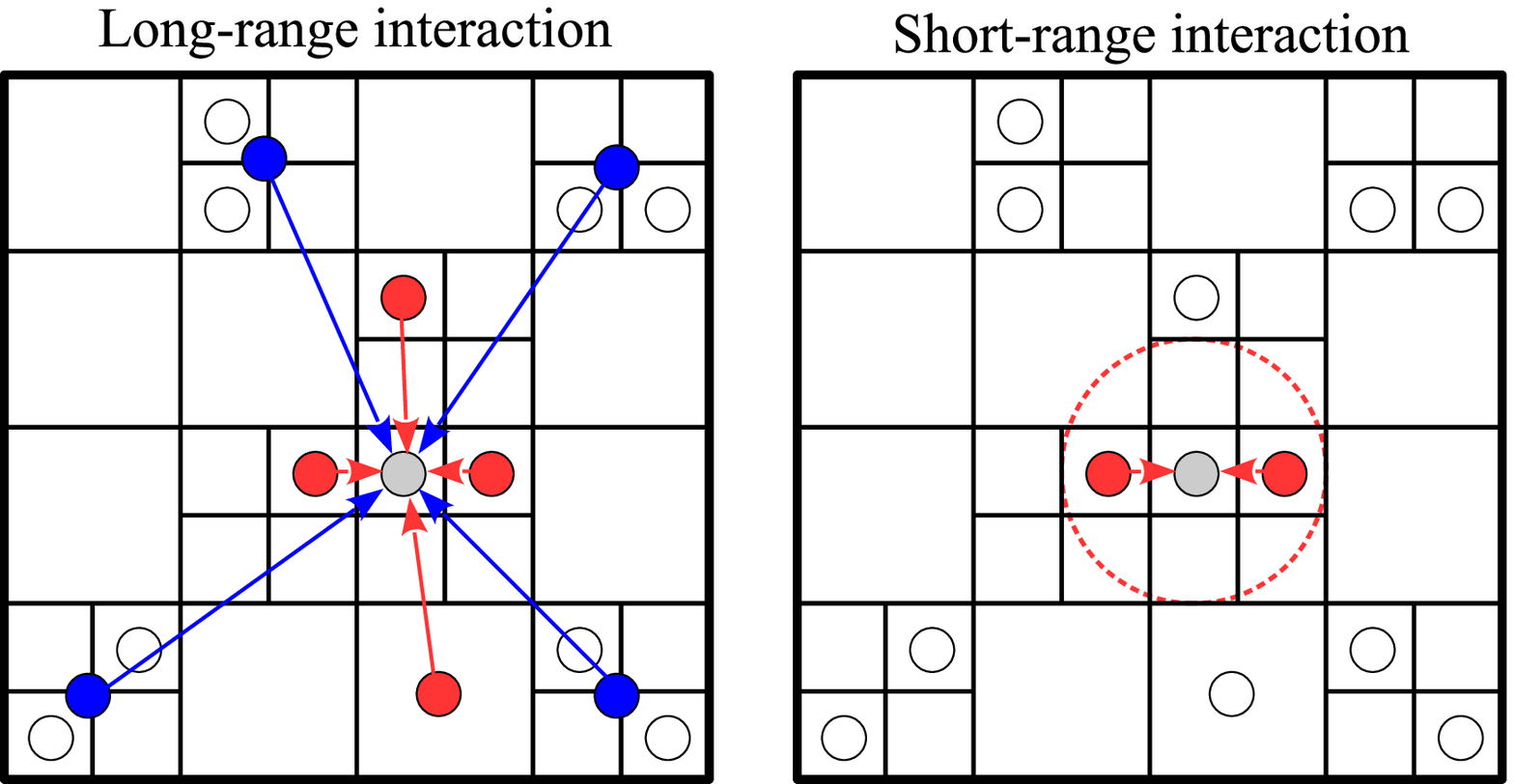}
  \end{center}
  \caption{Long-range interaction (left) and short-range interaction
    (right). Gray, red, and blue points are $i$-particles,
    $j$-particles, and superparticles, respectively.}
  \label{fig:forcetype}
\end{figure}

As stated earlier, FDPS performs the neighbor search if the interaction
is of short-range nature. If the long-range interaction is used,
currently FDPS uses the Barnes-Hut tree algorithm. In other words,
within FDPS, the distinction between the long-range and short-range
interactions is not a physical one but an operational one. If we want
to apply the treecode, it is a long-range interaction. Otherwise, it is
a short-range interaction.  Thus, we can use the simple tree
algorithm for pure $1/r$ gravity and the TreePM scheme
\citep{1995ApJS...98..355X, 2000ApJS..128..561B, 2002JApA...23..185B,
  2004NewA....9..111D, 2005MNRAS.364.1105S, 2005PASJ...57..849Y,
  2009PASJ...61.1319I, Ishiyama:2012:PAN:2388996.2389003} for the
periodic boundary.

Figure~\ref{fig:forcetype} illustrates the long-range and short-range
interactions and how they are calculated in FDPS.

For  long-range interactions, Barnes-Hut algorithm is used. Thus, the
interactions from a group of distant particles are replaced by that of
a superparticle, and  equation~(\ref{eq:geq}) is modified to 
%\begin{align}
\begin{eqnarray}
  \frac{d\myvec{u}_i}{dt} = \myvec{g}\left( \sum_j^{N_{\mathrm{J},i}}
  \myvec{f}(\myvec{u}_i,\myvec{u}_j) + \sum_{j'}^{N_{\mathrm{S},i}}
  \myvec{f'}(\myvec{u}_i,\myvec{u'}_{j'}), \myvec{u}_i
  \right), \label{eq:geqL}
%\end{align}
\end{eqnarray}
where $N_{\mathrm{J},i}$ and $N_{\mathrm{S},i}$ are, the number of
$j$-particles and superparticles for which we apply multipole-like
expansion, the vector $\myvec{u'}_{j'}$ is the physical quantity
vector of a superparticle, and the function $\myvec{f'}$ indicates the
interaction exerted on particle $i$ by the superparticle $j'$. In
simulations with a large number of particles $N$, $N_{\mathrm{J},i}$
and $N_{\mathrm{S},i}$ are many orders of magnitude smaller than $N$.
A user need to give functions to construct superparticles from
particles and to calculate the interaction from superparticles. Since
the most common use of the long-range interaction is for $1/r$
potential, FDPS includes standard implementation of these functions
for $1/r$ potential for  up to the quadrupole moment.

\subsection{An example --- gravitational \textit{N}-body problem}
\label{sec:samplecode}

In this section, we present a complete user code for the gravitational
$N$-body problem with the open boundary condition, to illustrate how a
user write an application program using FDPS.  The gravitational
interaction is handled as ``long-range'' type in FDPS. Therefore, we
need to provide the data type and interaction calculation functions
for superparticles. In order to keep the sample code short, we use the
center-of-mass approximation and use the same data class and
interaction function for real particles and superparticles.

For the gravitational
$N$-body problem, the physical quantity vector $\myvec{u}_i$ and interaction functions
$\myvec{f}$, $\myvec{f'}$, and $\myvec{g}$ are given by:
%\begin{align}
\begin{eqnarray}
  \myvec{u}_i &=& (\myvec{r}_i,
  \myvec{v}_i,m_i), \label{eq:PhysicalVectorNbody} \\
  \myvec{f} (\myvec{u}_i, \myvec{u}_j) &=& \frac{Gm_j \left(
    \myvec{r}_j - \myvec{r}_i \right)}{ \left( |\myvec{r}_j -
    \myvec{r}_i|^2 + \epsilon_i^2
    \right)^{3/2}}, \label{eq:ParticleParticleNbody} \\
  \myvec{f'} (\myvec{u}_i, \myvec{u'}_j) &=& \frac{Gm_j' \left(
    \myvec{r}_j - \myvec{r'}_i \right)}{ \left( |\myvec{r}_j -
    \myvec{r'}_i|^2 + \epsilon_i^2
    \right)^{3/2}}, \label{eq:ParticleSuperparticleNbody} \\
  \myvec{g}(\myvec{F},\myvec{u}_i)  &=& (\myvec{v}_i,\myvec{F},0), \\
  \myvec{F} &=& \sum_j
  \myvec{f}(\myvec{u}_i,\myvec{u}_j) + \sum_{j'}
  \myvec{f'}(\myvec{u}_i,\myvec{u'}_{j'}), \myvec{u}_i,
\label{eq:ConversionNbody}
%\end{align}
\end{eqnarray}
where $m_i$, $\myvec{r}_i$, $\myvec{v}_i$, and $\epsilon_i$ are, the
mass, position, velocity, and gravitational softening of particle $i$,
$m_j'$ and $\myvec{r'}_j$ are, the mass and position of a
superparticle $j$, and $G$ is the gravitational constant.  Note that
the shapes of the functions $\myvec{f}$ and $\myvec{f'}$ are the same.

Listing~\ref{code:samplecode} shows the complete code which can be
compiled and run, not only on a single-core machine but also
massively-parallel, distributed-memory machines such as the  K
computer. The total number of lines is 117.

%\begin{lstlisting}[label=code:samplecode,numbers=left,numbersep=5pt,frame=single,basicstyle=\ttfamily,caption=A sample code of $N$-body simulation]
%\begin{lstlisting}[label=code:samplecode]
%\begin{lstlisting}[label=code:samplecode,numbers=left,numbersep=5pt,frame=single,basicstyle=\ttfamily,caption=A sample code of $N$-body simulation]
\begin{lstlisting}[label=code:samplecode,numbers=left,numbersep=5pt,frame=single,basicstyle=\ttfamily]
#include <particle_simulator.hpp>
using namespace PS;

class Nbody{
public:
    F64    mass, eps;
    F64vec pos, vel, acc;
    F64vec getPos() const {return pos;}
    F64 getCharge() const {return mass;}
    void copyFromFP(const Nbody &in){ 
        mass = in.mass;
        pos  = in.pos;
        eps  = in.eps;
    }
    void copyFromForce(const Nbody &out) {
        acc = out.acc;
    }    
    void clear() {
        acc = 0.0;
    }
    void readAscii(FILE *fp) {
        fscanf(fp,
               "%lf%lf%lf%lf%lf%lf%lf%lf",
               &mass, &eps,
               &pos.x, &pos.y, &pos.z,
               &vel.x, &vel.y, &vel.z);
    }
    void predict(F64 dt) {
        vel += (0.5 * dt) * acc;
        pos += dt * vel;
    }
    void correct(F64 dt) {
        vel += (0.5 * dt) * acc;
    }
};

template<class TPJ>
struct CalcGrav{
    void operator () (const Nbody * ip,
                      const S32 ni,
                      const TPJ * jp,
                      const S32 nj,
                      Nbody * force) {
        for(S32 i=0; i<ni; i++){
            F64vec xi  = ip[i].pos;
            F64    ep2 = ip[i].eps
                * ip[i].eps;
            F64vec ai = 0.0;
            for(S32 j=0; j<nj;j++){
                F64vec xj = jp[j].pos;
                F64vec dr = xi - xj;
                F64 mj  = jp[j].mass;
                F64 dr2 = dr * dr + ep2;
                F64 dri = 1.0 / sqrt(dr2);                
                ai -= (dri * dri * dri
                       * mj) * dr;
            }
            force[i].acc += ai;
        }
    }
};

template<class Tpsys>
void predict(Tpsys &p,
             const F64 dt) {
    S32 n = p.getNumberOfParticleLocal();
    for(S32 i = 0; i < n; i++)
        p[i].predict(dt);
}

template<class Tpsys>
void correct(Tpsys &p,
             const F64 dt) {
    S32 n = p.getNumberOfParticleLocal();
    for(S32 i = 0; i < n; i++)
        p[i].correct(dt);
}

template<class TDI, class TPS, class TTFF>
void calcGravAllAndWriteBack(TDI &dinfo,
                             TPS &ptcl,
                             TTFF &tree) {
    dinfo.decomposeDomainAll(ptcl);
    ptcl.exchangeParticle(dinfo);    
    tree.calcForceAllAndWriteBack
        (CalcGrav<Nbody>(),
         CalcGrav<SPJMonopole>(),
         ptcl, dinfo);    
}

int main(int argc, char *argv[]) {
    PS::Initialize(argc, argv);                               
    F32 time  = 0.0;
    const F32 tend  = 10.0;
    const F32 dtime = 1.0 / 128.0;
    PS::DomainInfo dinfo;
    dinfo.initialize();
    PS::ParticleSystem<Nbody> ptcl;
    ptcl.initialize();
    PS::TreeForForceLong<Nbody, Nbody,
        Nbody>::Monopole grav;
    grav.initialize(0);
    ptcl.readParticleAscii(argv[1]);
    calcGravAllAndWriteBack(dinfo,
                            ptcl,
                            grav);
    while(time < tend) {
        predict(ptcl, dtime);        
        calcGravAllAndWriteBack(dinfo,
                                ptcl,
                                grav);
        correct(ptcl, dtime);        
        time += dtime;
    }
    PS::Finalize();
    return 0;
}
\end{lstlisting}

In the following we describe how this sample code works.  It consists
of four parts: the declaration to use FDPS (lines 1 and 2), the
definition of the particle (the vector $\myvec{u}_i$) (lines 4 to 35),
the definition of the gravitational force (the functions $\myvec{f}$
and $\myvec{f'}$) (lines 37 to 61), and the actual user program.  The
actual user program consists of a main routine and functions which
call library functions of FDPS (lines 63 to line 117). In the
following, we discuss each of them.

In order to use FDPS, the user program should
include the header file ``particle\_simulator.hpp''.
All functionalities of  the standard FDPS library are
implemented  as the header source library, since they are 
template libraries which need to receive particle class and
interaction functions.  FDPS data types and functions are in  the namespace
``PS''. In this sample program, we declare ``PS''  as the default
namespace to simplify the code. In this sample, however, we did not
omit ``PS'' for FDPS functions and class templates to show that they
come from FDPS.

FDPS defines several data types. \texttt{F32/F64} are data types of
32-bit and 64-bit floating points. \texttt{S32} is the data type of
32-bit signed integer.
\texttt{F64vec} is the class of a vector consisting of three 64-bit
floating points. This class provides several operators, such as the
addition, subtraction and the inner product indicated by ``$*$''.  It
is not necessary to use these data types in the user program, but some
of the functions the user should provide these data types for the
return value.

In the second part, the particle data type, ({\it i.e.} the vector
$\myvec{u}_i$) is defined as class \texttt{Nbody}. It has the
following member variables: \texttt{mass} ($m_i$), \texttt{eps}
($\epsilon_i$), \texttt{pos} ($\myvec{r}_i$), \texttt{vel}
($\myvec{v}_i$), and \texttt{acc} ($d\myvec{v}_i/dt$). Although the
member variable \texttt{acc} does not appear in
equation~(\ref{eq:PhysicalVectorNbody}) -- (\ref{eq:ConversionNbody}),
we need this variable to store the result of the gravitational force
calculation. A particle class for FDPS must provide public member
functions \texttt{getPos}, \texttt{copyFromFP},
\texttt{copyFromForce} and \texttt{clear} in these names, so that the internal functions
of FDPS can access the data within the particle class.  For the name
of the particle class itself and the names of the member variables, a
user can use whatever names allowed by the C++ syntax.  The member
functions \texttt{predict} and \texttt{correct} are used to integrate
the orbits of particles. These are not related to FDPS.  Since the
interaction is pure $1/r$ type, the construction method for
superparticles provided by FDPS can be used and they are not shown
here.

In the third part, the interaction functions $\myvec{f}$ and
$\myvec{f'}$ are defined. In this example, actually they are the same,
except for the class definition for $j$-particles. Therefore, this
argument is given as an argument with the template data
type \texttt{TPJ}, so that a single source code can be used to
generate two functions.  The interaction function used in FDPS should
have the following five arguments. The first argument \texttt{ip} is
the pointer to the array of $i$-particles which receive the
interaction. The second argument \texttt{ni} is the number of
$i$-particles. The third argument \texttt{jp} is the pointer to the
array of $j$-particles or superparticles which exert the
interaction. The fourth argument \texttt{nj} is the number of
$j$-particles or super-particles. The fifth argument \texttt{force} is
the pointer to the array of variables of a user-defined class to which
the calculated interactions on $i$-particles can be stored. In this
example, we used the particle class itself, but this can be another
class or a simple array.

In this example, the interaction function is a function object
declared as a \texttt{struct}, with the only member
function \texttt{operator ()}. FDPS can also accept a function pointer
for the interaction function, which would look a bit more familiar to
most readers.  In this example, the interaction is calculated through
a simple double loop. In order to achieve high efficiency, this part
should be replaced by a code optimized for specific architectures. In
other words, a user needs to optimize just this single function to
achieve very high efficiency.

In the fourth part, the main routine and user-defined functions are
defined.  In the following, we describe the main routine in detail,
and briefly discuss other functions. The main routine consists of the
following seven steps:
\begin{enumerate}
\item Initialize FDPS (line 92). \label{proc:init}
\item Set simulation time and timestep (lines 93 to 95). \label{proc:literal}
\item Create and initialize objects of FDPS classes (lines 96 to 102). \label{proc:construct}
\item Read in particle data from a file (line 103). \label{proc:input}
\item Calculate the gravitational forces on all the particles at the
  initial time (lines 104 to 106). \label{proc:calcinteraction}
\item Integrate the orbits of all the particles with Leap-Frog method
  (lines 107 to 114). \label{proc:integration}
\item Finish the use of  FDPS (line 115). \label{proc:fin}
\end{enumerate}

In the following, we describe  steps~\ref{proc:init},
\ref{proc:construct}, \ref{proc:input}, \ref{proc:calcinteraction},
and \ref{proc:fin}, and skip steps~\ref{proc:literal}
and \ref{proc:integration}.  In step~\ref{proc:literal}, we do not
call FDPS libraries.  Although we call FDPS libraries in
step~\ref{proc:integration}, the usage is the same as in
step~\ref{proc:calcinteraction}.

In step~\ref{proc:init}, the FDPS function \texttt{Initialize} is
called. In this function, MPI and OpenMP libraries are initialized. If
neither of them are used, this function does nothing.  All functions
of FDPS must be called between this function and the
function \texttt{Finalize}.

In step~\ref{proc:construct}, we create and initialize three objects
of the FDPS classes:
\begin{itemize}
\item \texttt{dinfo}: An object of class \texttt{DomainInfo}. It is
  used for domain decomposition.

\item \texttt{ptcl}: An object of class template \texttt{ParticleSystem}.
It takes the user-defined particle class (in this
example, \texttt{Nbody}) as the template argument. From the user
program, this object looks as an array of $i$-particles.

\item \texttt{grav}: An object of  data type \texttt{Monopole} defined in
class template \texttt{TreeForForceLong}. This object is used for the
calculation of long-range interaction using the tree algorithm.  It
receives three user-defined classes as template arguments: the class
to store the calculated interaction, the class for $i$-particles and
the class for $j$-particles. In this example, all these three classes
are the same as the original class of particles.  It is possible to
define classes with minimal data for these purposes and use them here,
in order to optimize the cache usage. The data type \texttt{Monopole}
indicates that the center-of-mass approximation is used for
superparticles.
\end{itemize}

In step~\ref{proc:input}, the data of particles are read from a file
into the object \texttt{ptcl}, using  FDPS
function \texttt{readParticleAscii}. In this function, a member
function of class \texttt{Nbody}, \texttt{readAscii}, is called. Note
that the user can write and use his/her own I/O functions. In this case,
\texttt{readParticleAscii} is unnecessary.

In step~\ref{proc:calcinteraction}, the forces on all particles are
calculated through the function \texttt{calcGravAllAndWriteBack},
which is defined in lines 79 to 89. In this function,
steps~\ref{proc:decompose} to
\ref{proc:interaction} in section~\ref{sec:view} are performed. In
other words, all of the actual work of FDPS libraries to calculate
interaction between particles takes place here. For
step~\ref{proc:decompose}, \texttt{decomposeDomainAll}, a member
function of class \texttt{DomainInfo} is called. This function takes
the object
\texttt{ptcl} as an argument to use the positions of particles to
determine the domain decomposition.  Step~\ref{proc:exchange} is
performed in \texttt{exchangeParticle}, a member function of
class \texttt{ParticleSystem}. This function takes the
object \texttt{dinfo} as an argument and redistributes particles among
MPI processes.  Steps \ref{proc:interactionexchange}
and \ref{proc:interaction} are performed
in \texttt{calcForceAllAndWriteBack}, a member function of
class \texttt{TreeForForceLong}. This function takes the user-defined
function object \texttt{CalcGrav} as the first and second arguments,
and calculates particle-particle and particle-superparticle
interactions using them.

In step~\ref{proc:fin}, the FDPS function \texttt{Finalize} is
called. It calls the \texttt{MPI\_Finalize} function.

In this section, we have described in detail how a user program
written using FDPS looks like. As we stated earlier, this program can
be compiled with or without parallelization using MPI and/or OpenMP,
without any change in the user program. The executable parallelized
with MPI is generated by using an appropriate compiler with MPI
support and a compile-time flag.  Thus, a user need not worry about
complicated bookkeeping necessary for parallelization using MPI.

In the next section, we describe how FDPS provides a generic
framework which takes care of parallelization
and bookkeeping for particle-based simulations.

\section{Implementation}
\label{sec:implementation}

In this section, we describe how the operations discussed in the
previous section are implemented in FDPS. In
section~\ref{sec:decomposition} we describe the domain decomposition
and particle exchange, and in section~\ref{sec:calculation}, the
calculation of interactions. Part of the contents in this scetion have
been published in \citet{2015FDPS}.

\subsection{Domain decomposition and particle exchange}
\label{sec:decomposition}

In this section, we describe how the domain decomposition and the
exchange of particles are implemented in FDPS. We used the
multisection method
\citep{2004PASJ...56..521M} with the so-called sampling
method \citep{Blackston:1997:HPE:509593.509597}. The multisection
method is a generalization of ORB (Orthogonal Recursive Bisection). In
ORB, as its name suggests, bisection is applied to each coordinate
axis recursively. In multisection method, division in one coordinate
is not to two domains but to an arbitrary number of domains. Since one
dimension can be divided to more than two sections, it is not
necessary to apply divisions many times. So we apply divisions only
once to each coordinate axis. A practical advantage of this method is
that the number of processors is not limited to powers of two.

\begin{figure}
  \begin{center}
    \includegraphics[width=8cm]{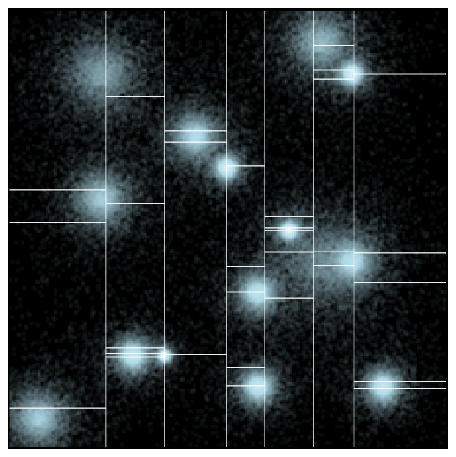}
  \end{center}
  \caption{Example of the domain decomposition. The division is $7
    \times 6$ in 2-dimension.}
  \label{fig:decomposition}
\end{figure}

Figure~\ref{fig:decomposition} illustrates an example of the
multisection method with $(n_x, n_y, n_z)=(7,6,1)$. We can see that
the size and shape of subdomains show large variation. By allowing
this variation, FDPS achieves quite good load balance and high
scalability. Note that $n=n_x n_y n_z$ is the number of MPI
processes. By default, values of $n_x$, $n_y$, and $n_z$ are chosen so
that they are integers close to $n^{1/3}$. For figure
~\ref{fig:decomposition}, we force the numbers used to make a
two-dimensional decomposition.

In the sampling method, first each process performs random sampling of
particles under it, and sends them to the process with rank 0
(``root'' process hereafter). Then the root process calculates the
division so that sample particles are equally divided over all
processes, and broadcasts the geometry of domains to all other
processes. In order to achieve good load balance, sampling frequency
should be changed according to the calculation cost per particle
\citep{2009PASJ...61.1319I}.

The sampling method works fine, if the number of particles per process
is significantly larger than the number of process. This is, however,
not the case for runs with a large number of nodes.  When the number
of particles per process is not much larger than the number of
processes, the total number of sample particles which the root process
needs to handle exceeds the number of particles per process itself,
and thus calculation time of domain decomposition in the root process
becomes visible.

In order to reduce the calculation time, we also parallelized the
domain decomposition, currently in the direction of $x$ axis only. The
basic idea is that each node sends the sample particles not to the
root process of the all MPI processes but to the processes with index
$(i,0,0)$. Then processes $(i,0,0)$ sort the sample particles and
exchange the number of sample particles they received. Using these two
pieces of information, each of $(i,0,0)$ processes can determine all
domain boundaries inside its current domain in the $x$ direction. Thus,
they can determine which sample particles should be sent to
where. After the exchange of sample particles, each of $(i,0,0)$
processes can determine the decompositions in $y$ and $z$ directions.

A naive implementation of the above algorithm requires ``global''
sorting of sample particles over all of $(i,0,0)$ processes. In order
to simplify this part, before each process sends the sample particles
to  $(i,0,0)$ processes, they exchange their samples with other
processes with the same location in $y$ and $z$ process coordinates, so
that they have sample particles in the current domain decomposition in
the x direction. As a result, particles sent to $(i,0,0)$ processes
are already sorted at the level of domains decomposition in $x$
direction, and we need only the  sorting within each of $(i,0,0)$
processes to obtain the globally sorted particles.

Thus, our implementation of parallelized domain decomposition
is as follows:

\begin{enumerate}
\item
Each process samples particles randomly from its own particles. In
order to achieve an optimal load balance, the sampling rate of
particles is changed so that it is proportional to the CPU time per
particle spent on that process \citep{2009PASJ...61.1319I}. FDPS
provides several options including this optimal
balance. \label{prcoc:sampling}

\item
Each process exchanges the sample particles according to the current
domain boundary in the $x$ direction with the process with the same y
and z indices, so that they have sample particles in the current
domain decomposition in the $x$ direction.

\label{proc:commx}

\item
Each process with index $(i,y,z)$ sends the sample particles to the
process with index $(i,0,0)$, in other words, the root processes in
each of $y$-$z$ planes collects subsamples.

\label{proc:gatherx}

\item
Each root process sorts the sample particles in the $x$ direction. Now,
the sample particles are sorted globally in the $x$ direction.

\item
Each root process sends the number of the sample particles to the
other root processes and determines the global rank of the sample
particles.

\item
Determine the $x$ coordinate of new domains by dividing all sample
particles into $n_x$ subsets with equal number of sample particles.
\label{proc:determinex}

\item
Each root process exchanges sample particles with other root
processes, so that they have the sample particles in new domain in the
$x$ direction.

\item
Each root process determines the $y$ and $z$ coordinates of new domains.
\label{proc:detyz}

\item
Each root process broadcasts the geometries of new domains to all
other processes.
\label{proc:broadcasting}

\end{enumerate}

It is also possible to parallelize  the determination of  subdomains in
step \ref{proc:detyz}, but even for the full-node runs on K computer
we found the current parallelization is sufficient.

For particle exchange and also for interaction information exchange,
we use {\tt MPI\_Alltoall} to exchange the length of the data and {\tt
MPI\_Isend} and {\tt MPI\_Irecv} to actually exchange the data. At
least on K computer, we found that the performance of vendor-provided
{\tt MPI\_Alltoall} is not optimal for short messages. We implemented
a hand-crafted version in which the messages sent to the same relay
points are combined in order to reduce the total number of messages.

After the domain decomposition is done and the result is broadcasted
to all processes, they exchange particles so that each of them has
particles in its domain. Since each process has the complete
information of the domain decomposition, this part is pretty
straightforward to implement. Each process looks at each of its
particles, and determines if that particle is still in its domain.  If
not, the process determines to which process that particle should be
sent. After the destinations of all particles are determined, each
process sends them out, using {\tt MPI\_Isend} and {\tt MPI\_Irecv}
functions.

\subsection{Interaction calculation}
\label{sec:calculation}

In this section, we describe the implementations of interaction
information exchange and actual interaction calculation.
In the interaction information exchange step,
each process sends the  data required by other nodes. In the
interaction calculation step, actual interaction calculation is done
using the received data.
For both steps, the Barnes-Hut octree structure is used, for both of
long- and short-range interactions.

First, each process constructs the tree of its local particles. Then
this tree is used to determine the data to be sent to other
processes. For the long-range interaction, the procedure is the same
as that for usual tree traversal\citep{1986Natur.324..446B,
1990JCoPh..87..161B}.  The tree traversal is used also for short-range
interactions.  FDPS can currently handle four different types of the
cutoff length for the short-range interactions: fixed, $j$-dependent,
$i$-dependent and symmetric.  For $i$-dependent and symmetric cutoffs,
the tree traversal should be done twice.

Using the received data, each process performs the force
calculation. To do so, it first constructs the tree of all data
received and local particles, and then uses it to calculate  the
interaction on local particles.

\section{Performance of applications developed using FDPS}
\label{sec:performance}

In this section, we present the performance of three astrophysical
applications developped using FDPS. One is the pure gravity code with
open boundary applied to disk galaxy simulation. The second one is
again pure gravity application but with periodic boundary applied to
cosmological simulation. The third one is gravity + SPH calculation
applied to the giant impact (GI) simulation.  For the performance
measurement, we used two systems. One is K computer of RIKEN AICS, and
the other is Cray XC30 of CfCA, National Astronomical Observatory of
Japan. K computer consists of 82,944 Fujitsu SPARC64 VIIIfx
processors, each with eight cores. The theoretical peak performance of
one core is 16 Gflops, for both of single- and double-precision
operations. Cray XC30 of CfCA consists of 1060 nodes, or 2120 Intel
Xeon E5-2690v3 processors (12 cores, 2.6GHz). The theoretical peak
performance of one core is 83.2 and 41.6 Gflops for single- and
double-precision operations, respectively.  In all runs on K computer,
we use the hybrid MPI-OpenMP mode of FDPS, in which one MPI process is
assigned to one node. On the other hand, for XC30, we use the flat MPI
mode of FDPS. The source code is the same except for that for the
interaction calculation functions. The interaction calculation part
was written to take full advantage of the SIMD instruction set of the
target architecture, and thus written specifically for SPARC64 VIIIfx
(HPC-ACE instruction set) and Xeon E5 v3 (AVX2 instruction set).

\label{sec:measuredperformance}
\subsection{Disk galaxy simulation}
\label{sec:diskgalaxy}
In this section, we discuss the performance and scalability of a
gravitational $N$-body simulation code implemented using FDPS. Some
results in this scetion have been published in \citet{2015FDPS}. The
code is essentially the same as the sample code described in
section~\ref{sec:samplecode}, except for the following two differences
in the user code for the calculation of the interaction. First, to
improve the accuracy, we used the expansion up to the quadrupole
moment, instead of the monopole-only one used in the sample
code. Second, we used the highly optimized kernel developed using SIMD
builtin functions, instead of the simple one in the sample code.

We apply this code for the simulation of the Milky Way-like galaxy,
which consists of a bulge, a disk, and a dark matter halo. For
examples of recent large-scale simulations,
see \citet{2011ApJ...730..109F}
and \citet{Bedorf:2014:PGT:2683593.2683600}.

The initial condition is the Milky Way model, which is the same as
that in \citet{Bedorf:2014:PGT:2683593.2683600}. The mass of the bulge
is $4.6 \times 10^9 M_\odot$, and it has a spherically-symmetric
density profile of the Hernquist model \citep{1990ApJ...356..359H}
with the half-mass radius of $0.5$~kpc. The disk is an axisymmetric
exponential disk with the scale radius of $3$~kpc, the scale height of
$200$~pc and the mass $5.0 \times 10^{10}M_\odot$. The dark halo has
an Navarro-Frenk-White (NFW) density
profile \citep{1996ApJ...462..563N} with the half-mass radius of
$40$~kpc and the mass of $6.0 \times 10^{11} M_\odot$. In order to
realize the Milky Way model, we used
GalacticICS \citep{2005ApJ...631..838W}. For all simulations in this
section, we adopt $\theta=0.4$ for the opening angle for the tree
algorithm. We set the average number of particles sampled for the
domain decomposition to 500.

\begin{figure}
  \begin{center}
    \includegraphics[width=8cm]{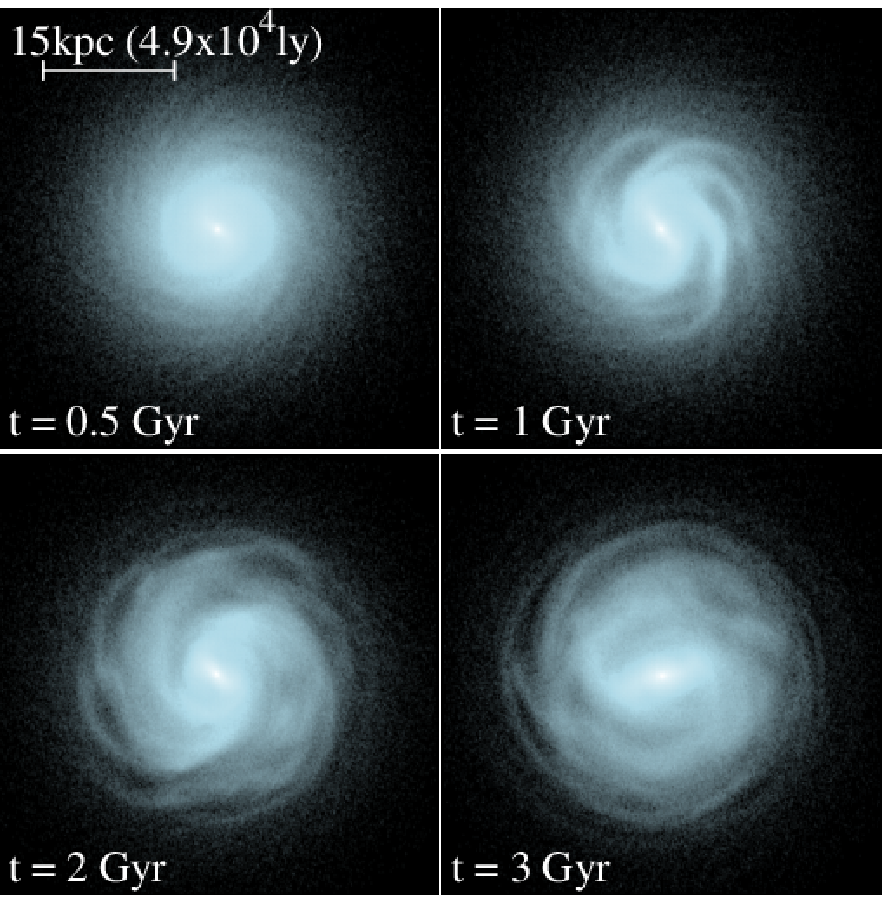}
  \end{center}
  \caption{
  Face-on surface density maps of the bulge and disk.
  }
  \label{fig:evolutiondisk}
\end{figure}

Figure~\ref{fig:evolutiondisk} illustrates the time evolution of the
bulge and disk in the run with $512$ nodes on the K computer. The disk
is initially axisymmetric. We can see that spiral structure develops
(0.5 and 1 Gyrs) and a central bar follows the spiral (1 Gyrs and
later). As the bar grows, the two-arm structure becomes more visible
(3 Gyrs).

\begin{figure}
  \begin{center}
    \includegraphics[width=8cm]{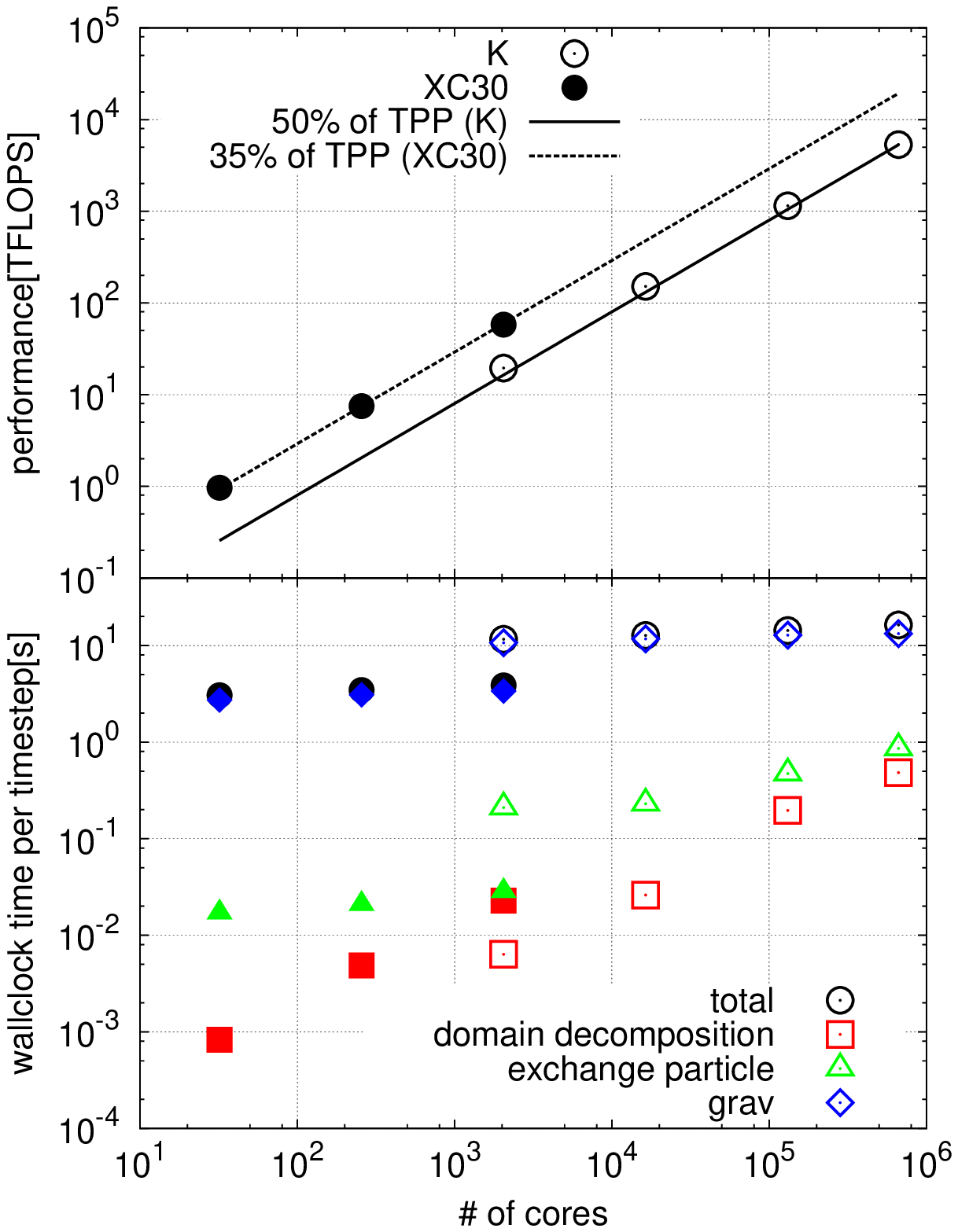}
  \end{center}
  \caption{
  
    Weak-scaling performance of the gravitational $N$ body code. The
    speed of the floating-point operation (top) and wallclock time per
    one timestep (bottom) are plotted as functions of the number of
    cores. Open and filled symbols indicate the performances of K
    computer and cray XC30, respectively. In the top panel, the solid
    line indicates 50\% of the theoretical peak performance of K
    computer and the dotted line indicates 35\% of the theoretical
    peak performance of XC30. In the bottom panel, time spent for the
    interaction calculation (diamond), the domain decomposition
    (square) the exchange particles (triangle) are also shown.
    
    } \label{fig:disk_weak}
\end{figure}

Figure~\ref{fig:disk_weak} shows the measured weak-scaling
performance. We fixed the number of particles per core to 266,367 and
measured the performance for the number of cores in the range of 4096
to 663,552 on the K computer, and in the range of 32 to 2048 on
XC30. We can see that the measured efficiency and scalability are both
very good. The efficiency is more than 50\% for the entire range of
cores on the K computer. The efficiency of XC30 is a bit worse than
that of the K computer. This difference comes from the difference of
two processors. The Fujitsu processor showed higher efficiency, while
the Intel processor has 5.2 times higher peak performance per core. We
can see that the time for domain decomposition increase as we increase
the number of cores. The slope is around 2/3 as can be expected from
our current algorithm discussed in section \ref{sec:decomposition}.

\begin{figure}
  \begin{center}
    \includegraphics[width=8cm]{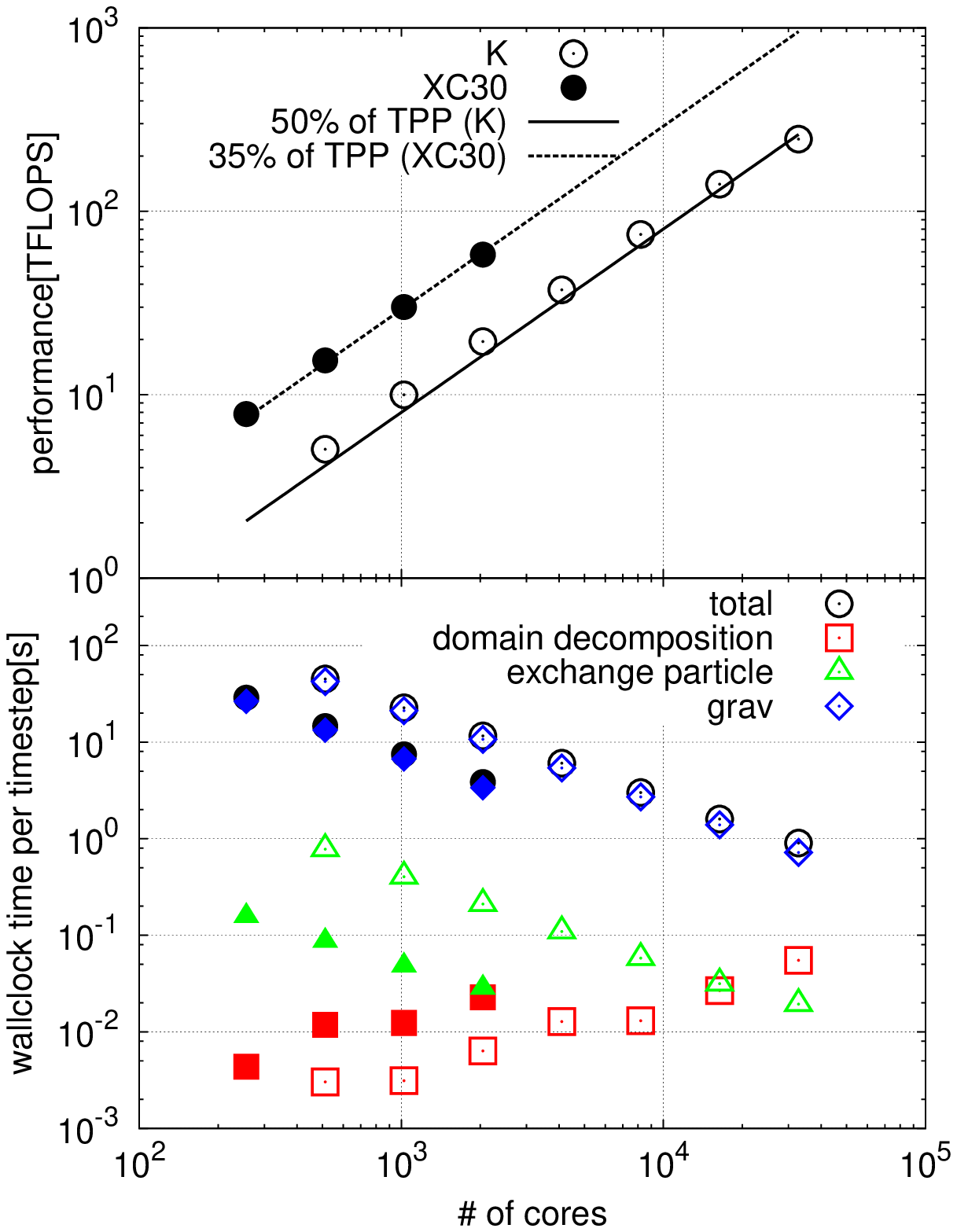}
  \end{center}
  \caption{
  
The same figure as figure \ref{fig:disk_weak} but for the
strong-scaling performance for 550 million particles

    }
  \label{fig:disk_strong}
\end{figure}

Figure~\ref{fig:disk_strong} shows the measured strong-scaling
performance. We fixed the total number of particles to $550$ million
and measured the performance for 512 to 32768 cores on K computer and
256 to 2048 cores on XC30. We can also see the measured efficiency and
scalability are both very good, for the strong-scaling performance.

\citet{Bedorf:2014:PGT:2683593.2683600}
reported the wallclock time of 4 seconds for their 27-billion particle
simulation on the Titan system with 2048 NVIDIA Tesla K20X, with the
theoretical peak performance of 8PF (single precision, since the
single precision was used for the interaction calculation). This
corresponds to 0.8 billion particles per second per petaflops. Our
code on K computer requires 15 seconds on 16384 nodes (2PF theoretical
peak), resulting in 1 billion particles per second per
petaflops. Therefore, we can conclude that our FDPS code achieved the
performance slightly better than one of the best codes specialized to
gravitational $N$-body problem.

\subsection{Cosmological simulation}

In this section, we discuss the performance of a cosmological
simulation code implemented using FDPS. We implemented TreePM (Tree
Particle-Mesh) method and measured the performance on XC30. Our TreePM
code is based on the code developed by K. Yoshikawa. The Particle-Mesh
part of the code was developed by
\citet{Ishiyama:2012:PAN:2388996.2389003} and this code is included in
the FDPS package as an external module.

We initially place particles uniformly in a cube and gave them zero
velocity. For the calculation of the tree force , we used a monopole
only kernel with cutoff. The cutoff length of the force is three times
larger than the width of the mesh. We set $\theta$ to 0.5. For the
calculation of the mesh force, the mass density is assigned to each of
the grid points, using the triangular shaped cloud scheme and the
density profile we used is the S2 profile \citep{hockney1988computer}.

\begin{figure}
  \begin{center}
    \includegraphics[width=8cm]{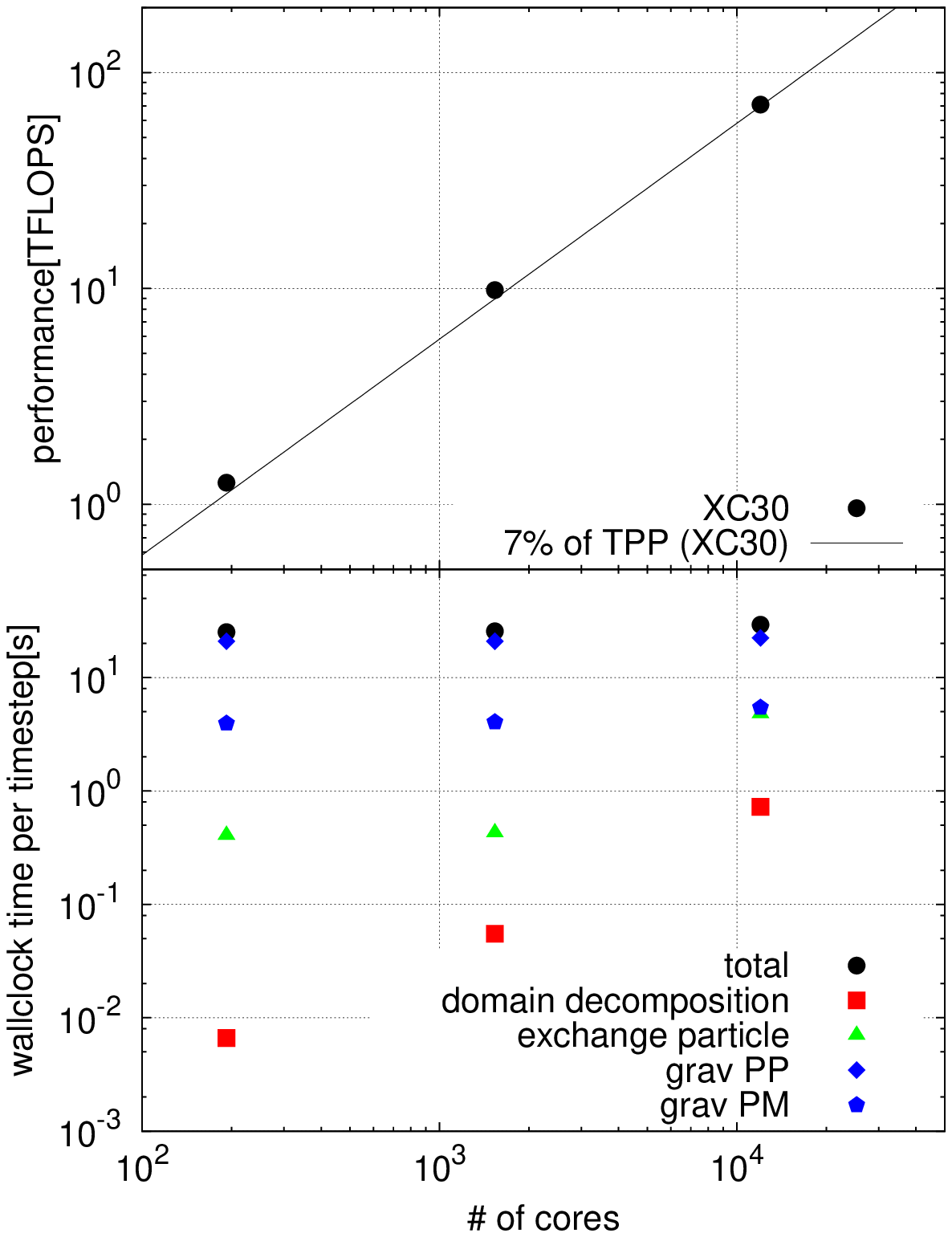}
  \end{center}
  \caption{

    Weak-scaling performance of the TreePM code. The speed of the
    floating-point operation (top) and wallclock time per one timestep
    (bottom) are plotted as functions of the number of cores. In the
    top panel, the solid line indicates 7\% of the theoretical peak
    performance of XC30. In the bottom panel, time spent for the
    Particle-Particle interaction calculation (diamond), the
    Particle-Mesh interaction (pentagon), the domain decomposition
    (square) and the exchange particles (triangle) are also shown.
    
  }
  \label{fig:cosmo_weak}
\end{figure}

\begin{figure}
  \begin{center}
    \includegraphics[width=8cm]{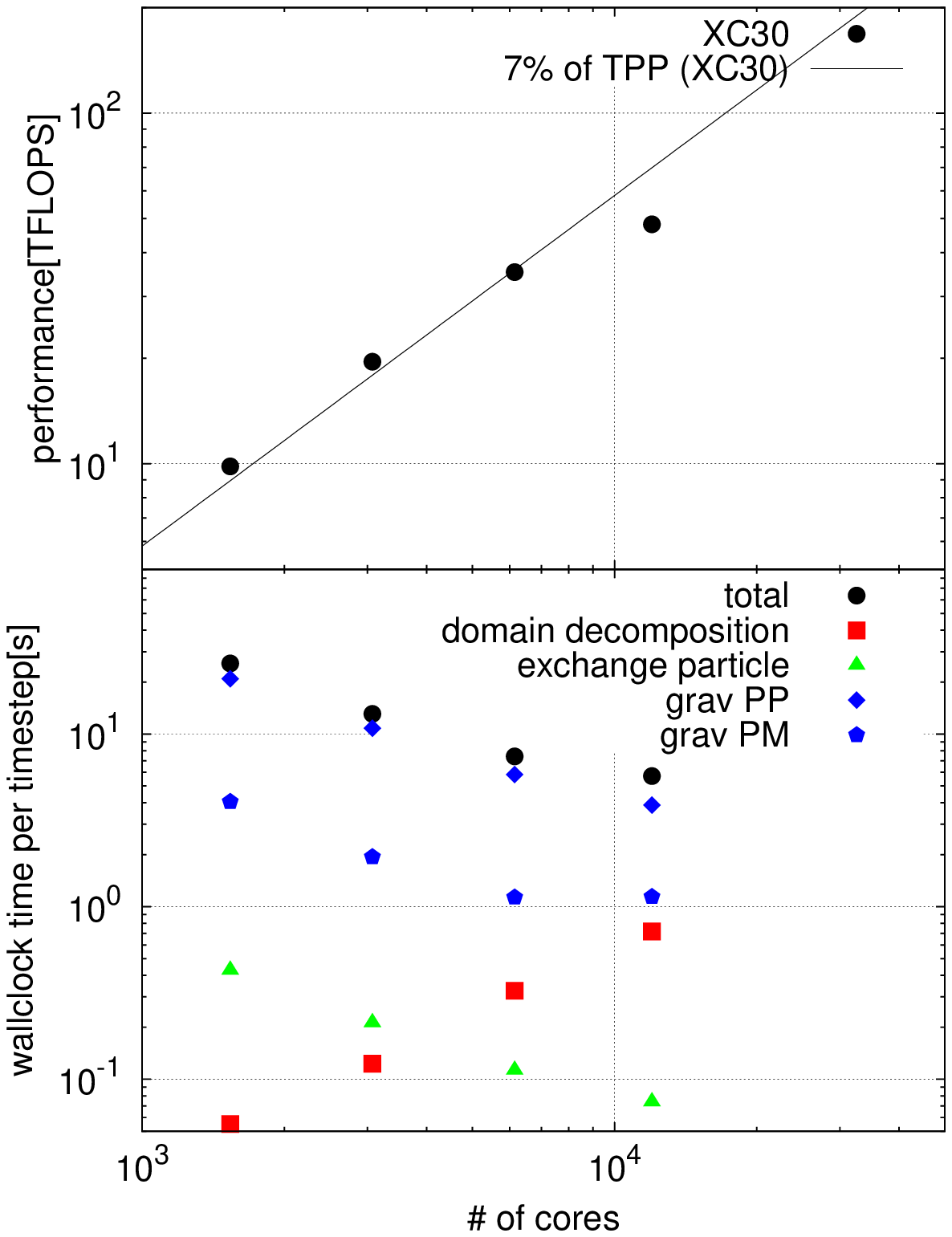}
  \end{center}
  \caption{

    The same as figure \ref{fig:cosmo_weak} but for the strong-scaling
    performance. In this case, the number of particles is $2048^3$.

  }
  \label{fig:cosmo_strong}
\end{figure}

Figures \ref{fig:cosmo_weak} and \ref{fig:cosmo_strong} show the weak
and strong scaling performance, respectively. For the weak-scaling
measurement, we fixed the number of particles per process to 5.73
million and measured the performance for the number of cores in the
range of 192 to 12000 on XC30. For the strong-scaling measurements, we
fixed the total number of particles to $2048^3$ and measured the
performance for the number of cores in the range of 1536 to 12000 on
XC30. We can see that the time for the calculation of the tree force
is dominant and both of the weak and strong scalings are good except
for the very large number of cores (12000) for the strong scaling
measurement. One reason is that the scalability of the calculation of
the mesh force is not very good. Another reason is that the time for
the domain decomposition grows linearly for large number of cores,
because we did not use parallelized domain decomposition here. The
efficiency is 7\% of the theoretical peak performance. It is rather
low compared to that for the disk galaxy simulations in
section \ref{sec:diskgalaxy}. The main reason is that we use a lookup
table for the force calculation. If we evaluate the force without the
lookup table, the nominal efficiency would be much better, but the
total time would be longer.

\subsection{Giant impact simulation}

\label{sec:sph}

In this section, we discuss the performance of an SPH simulation code
with self-gravity implemented using FDPS.  Some results in this
scetion have been published in \citet{2015FDPS}.  The test problem
used is the simulation of GI. The GI
hypothesis \citep{1975Icar...24..504H, 1976LPI.....7..120C} is one of
the most popular scenarios for the formation of the Moon. The
hypothesis is as follows. About 5 billion years ago, a Mars-sized
object (hereafter, the impactor) collided with the proto-Earth
(hereafter, the target). A large amount of debris was scattered, which
first formed the debris disk and eventually the Moon. Many researchers
have performed simulations of GI, using the SPH method
\citep{1986Icar...66..515B, 2013Icar..222..200C, 2014NatGe...7..564A}.

For the gravity, we used monopole-only kernel with $\theta=0.5$. We
adopt the standard SPH scheme
\citep{1992ARA&A..30..543M, 2009NewAR..53...78R, 2010ARA&A..48..391S}
for the hydro part. Artificial viscosity is used to handle shocks
\citep{1997JCoPh.136..298M}, and 
the standard Balsara switch is used to reduce the shear viscosity
\citep{1995JCoPh.121..357B}. A kernel function we used is the Wendland $C^6$ and the cutoff radius
is 4.2 times larger than the local mean inter-particle distance. In
other words, each particle interact with about 300 particles. This
neighbor number is the appropriate for this kernel to avoid the
pairing instability \citep{2012MNRAS.425.1068D}.

Assuming that the target and impactor consist of granite, we adopt
equation of state of granite \citep{1986Icar...66..515B} for the
particles. For the initial condition, we assume the parabolic orbit
with the initial angular momentum 1.21 times of the current Earth-Moon
system.

\begin{figure}
  \begin{center}
    \includegraphics[width=8cm]{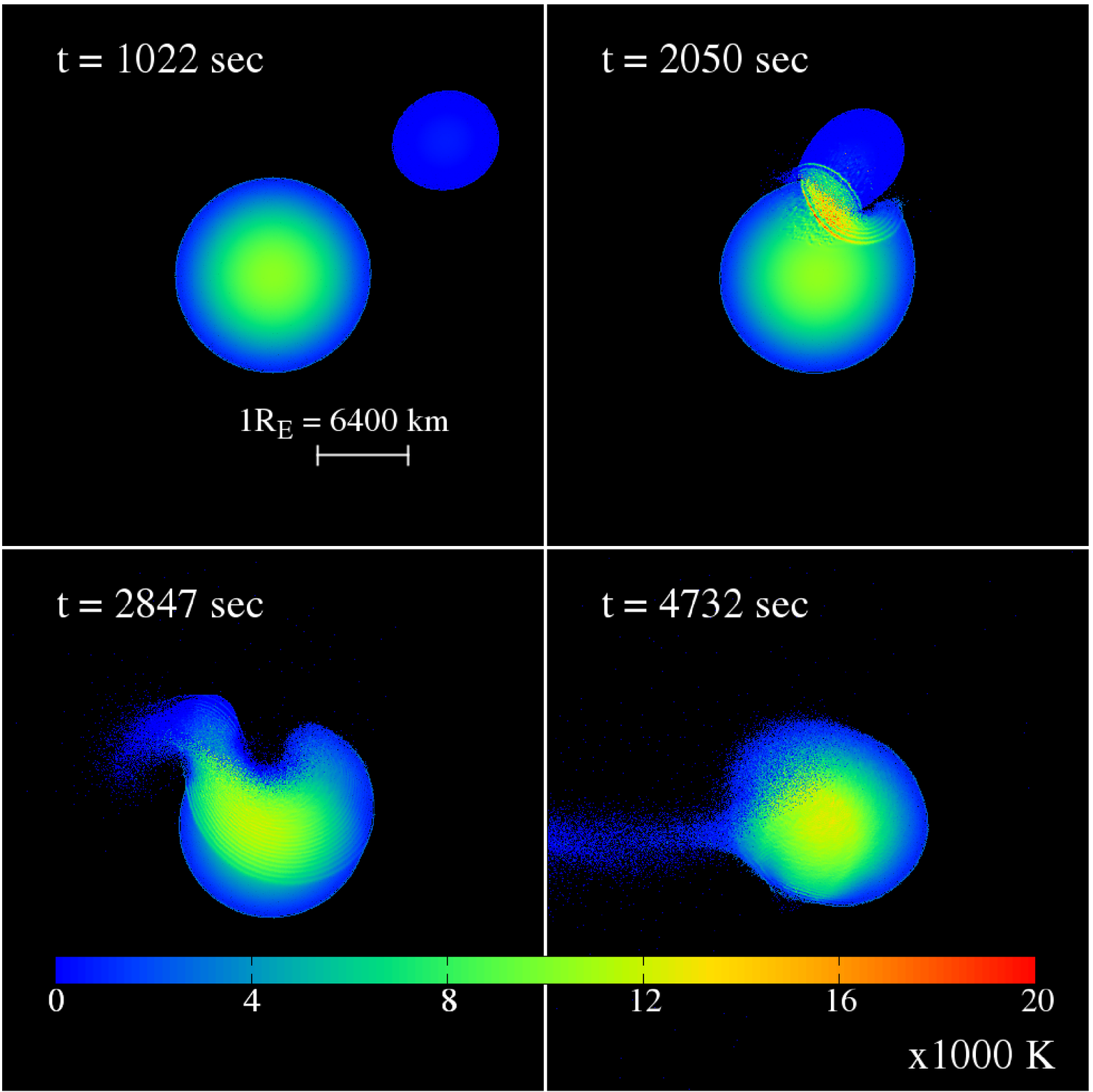}
  \end{center}
  \caption{Temperature maps of the target and impactor in the run with
  $9.9$ million particles at four different epochs. }
  \label{fig:evolutionGI}
\end{figure}

Figure~\ref{fig:evolutionGI} shows the time evolution of the target
and impactor for a run with 9.9 million particles. We can see that the
shocks are formed just after the moment of impact in both the target
and impactor ($t=2050$ sec). The shock propagates in the target, while
the impactor is completely disrupted ($t=2847$ sec) and debris are
ejected. A part of the debris falls back to the target, while the rest
will eventually form the disk and the Moon. So far, the resolution
used in the published papers have been much lower. We plan to use this
code to improve the accuracy of the GI simulations.

\begin{figure}
  \begin{center}
    \includegraphics[width=8cm]{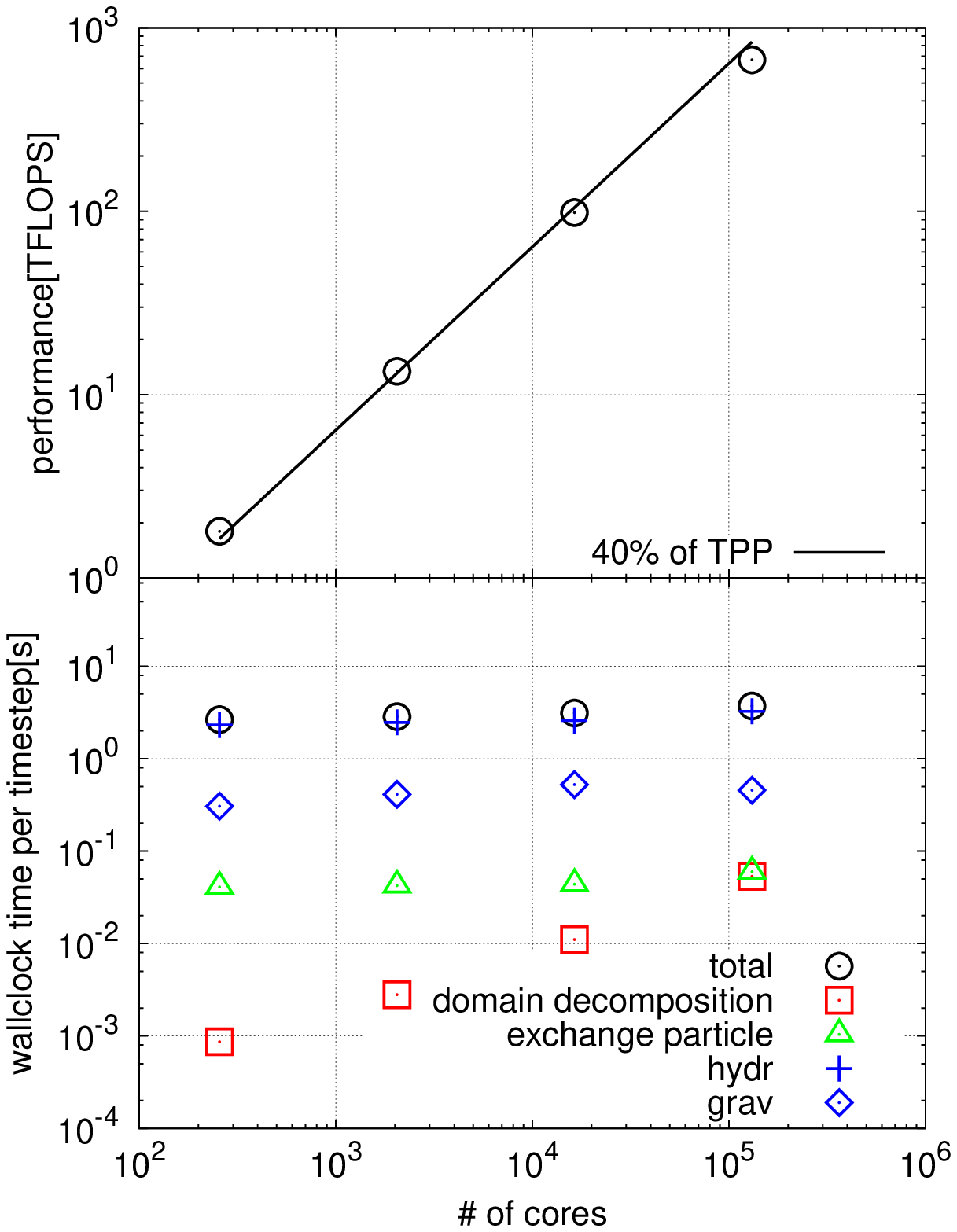}
  \end{center}
  \caption{

    Weak-scaling performance of the SPH code. The speed of the
    floating-point operation (top) and wallclock time per one timestep
    (bottom) are plotted as functions of the number of cores. In the
    top panel, the solid line indicates 40\% of the theoretical peak
    performance of K computer. In the bottom panel, time spent for the
    hydrodynamics calculation (cross), the gravity calculation
    (diamond), the domain decomposition (square) the exchange
    particles (triangle) are also shown.
}
  \label{fig:gi_weak}
\end{figure}

\begin{figure}
  \begin{center}
    \includegraphics[width=8cm]{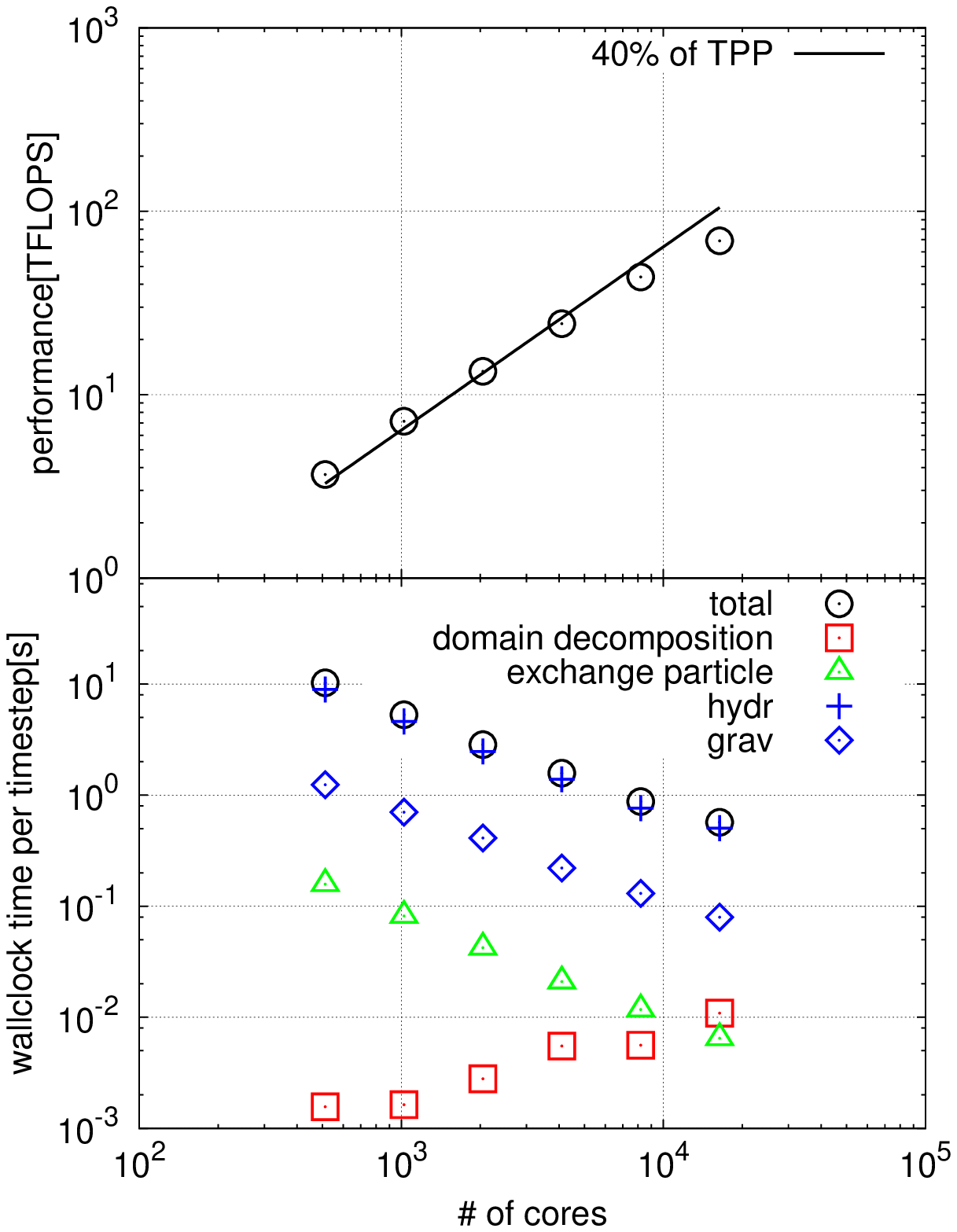}
  \end{center}
  \caption{

The same as figure \ref{fig:gi_weak} but for the strong-scaling
performance for $39$ million particles.

}
  \label{fig:gi_strong}
\end{figure}

Figure~\ref{fig:gi_weak} and \ref{fig:gi_strong} show the measured
weak and strong scaling performance. For the weak-scaling measurement,
we fixed the number of particles per core to 20,000 and measured the
performance for the number of cores in the range of 256 to 131,072 on
the K computer. On the other hand, for the strong-scaling measurement,
we fixed the total number of particles to $39$ million and measured
the performance for the number of cores in the range of 512 to 16,384
on K computer. We can see that the performance is good even for very
large number of cores. The efficiency is about 40\% of the theoretical
peak performance. The hydro part consumes more time than the gravity
part does, mainly because the particle-particle interaction is more
complicated.

\section{Performance model}
\label{sec:performancemodel}

In this section, we present the performance model of applications
implemented using FDPS. As described in section \ref{sec:user}, the
calculation of a typical application written using FDPS proceeds in
the following steps

\begin{enumerate}
  \item Update the domain decomposition and exchange particles
    accordingly (not in every timestep).
  \item  Construct the local tree structure and exchange particles and
    superparticles necessary for interaction calculation.
  \item Construct the ``global'' tree.
  \item Perform the interaction calculation.
  \item Update the physical quantities of particles using the
    calculated interactions.
\end{enumerate}

In the case of complex applications which require more than one
interaction calculations, each of the above steps, except for the
domain decomposition, may be executed more than one time per one
timestep.

For a simple application, thus, the total wallclock time per one timestep should
be expressed as
\begin{equation}
  \label{eq:totalcost}
  T_{\rm step} =  T_{\rm dc}/n_{\rm dc}
               + T_{\rm lt}
               + T_{\rm exch}
               + T_{\rm icalc}
               + T_{\rm misc},
\end{equation}
where   $T_{\rm dc}$, $T_{\rm lt}$, $T_{\rm exch}$,
$T_{\rm icalc}$, and $T_{\rm misc}$ are the times for
domain composition and particle exchange, local tree construction,
exchange of particles and superparticles for interaction calculation,
interaction calculation, and other calculations such as particle
update, respectively. The term $n_{\rm dc}$ is the interval at which
the domain decomposition is performed.

In the following, we first construct the model for the communication
time. Then we construct models for each term of the right hand side of
equation~\ref{eq:totalcost}, and finally we compare the model with the
actual measurement presented in section
\ref{sec:total_time}.

\subsection{Communication model}
\label{sec:comm_model}

What ultimately determines the efficiency of a calculation performed
on a large-scale parallel machine is the communication overhead. Thus,
it is very important to understand what types of communication would
take what amount of time on actual hardware. In this section, we
summarize the characteristics of the communication performance of K
computer.

In FDPS, almost all communications are through the use of collective
communications, such as {\tt MPI\_Allreduce}, {\tt MPI\_Alltoall}, and
{\tt MPI\_Alltoallv}. However, measurement of the performance of these
routines for uniform message length is not enough, since the amount of
data to be transferred between processes generally depends on the
physical distance between domains assigned to those
processes. Therefore, we first present the timing results for simple
point-to-point communication, and then for collective communications.

\begin{figure}
  \begin{center}
    \includegraphics[width=8cm]{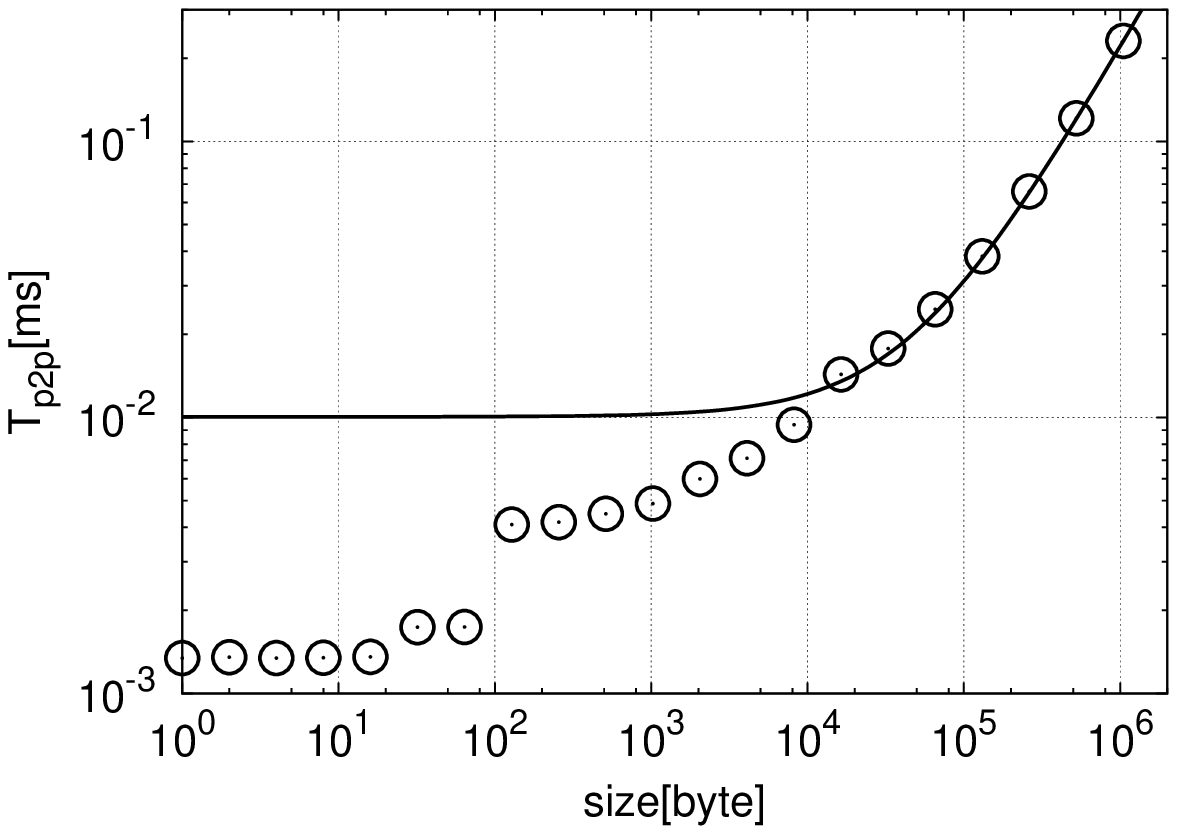}
  \end{center}
  \caption{
  
  Elapsed time for point-to-point communication as a function of size
  of message measured on K computer.

}
  \label{fig:pingpong}
\end{figure}

Figure \ref{fig:pingpong} shows the elapsed time as the function of
the message length, for point-to-point communication between
``neighboring'' processes. In the case of K computer, we used
three-dimensional node allocation, so that ``neighboring'' processes
are actually close to each other in its torus network.

We can see that the elapsed time can be fitted reasonably well as
\begin{equation}
\label{eq:Tp2p}
  T_{\rm p2p} = T_{\rm p2p,startup} + n_{\rm word}T_{\rm p2p,word},
\end{equation}
where $T_{\rm p2p,startup}$ is the startup time which is independent
of the message length and $T_{\rm p2p,word}$ is the time to transfer
one byte of message. Here, $n_{\rm word}$ is the length of the message
in units of bytes. On K computer, $T_{\rm p2p,startup}$ is 0.0101 ms
and $T_{\rm p2p,word}$ is $2.11 \times 10^{-7}$ ms per byte. For a
short message, there is a rather big discrepancy between the analytic
model and measured points, because for short messages K computer used
several different algorithms.

\begin{figure}
  \begin{center} \includegraphics[width=8cm]{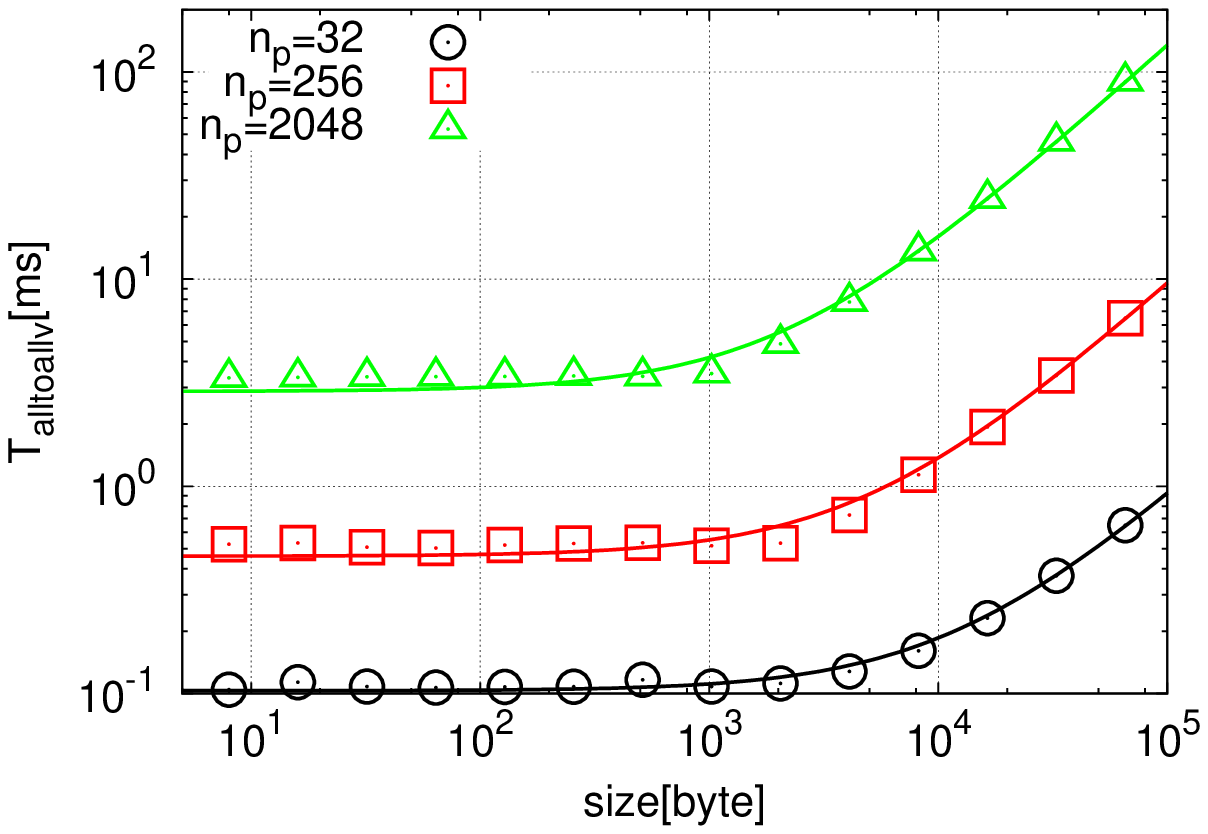}
  \end{center}
  \caption{
  
    Elapsed time of {\tt MPI\_Alltoallv} as a function of message size
    measured on K computer.

}
\label{fig:group_comm_a2a}
\end{figure}

Figure \ref{fig:group_comm_a2a} shows the elapsed times for {\tt
MPI\_Alltoallv}. The number of processes $n_p$ is 32 to 2048. They are
again modeled by the simple form
\begin{equation}
\label{eq:wtime_group_comm}
  T_{\rm alltoallv} = T_{\rm alltoallv,\rm startup} + n_{\rm word}T_{\rm alltoallv,\rm word}, 
\end{equation}
where $T_{\rm \rm alltoallv,\rm startup}$ is the startup time and
$T_{\rm alltoallv,\rm word}$ is the time to transfer one byte of message.
We list these values in table \ref{table:alltoallv}.

\begin{table}
\caption{Time coefficients in equation (\ref{eq:wtime_group_comm})}
\begin{tabular}{|l|l|l|l|} \hline
                  & $n_p=32$ & $n_p=256$ & $n_p=2048$ \\ \hline
$T_{\rm alltoallv,startup}$ [ms] & 0.103 & 0.460 & 2.87 \\
$T_{\rm alltoallv,word}$ [ms/byte] & $8.25\times 10^{-6}$  & $9.13\times 10^{-5}$ & $1.32\times 10^{-3}$  \\ \hline
\end{tabular}
\label{table:alltoallv}
\end{table}

\begin{figure}
  \begin{center}
  \includegraphics[width=8cm]{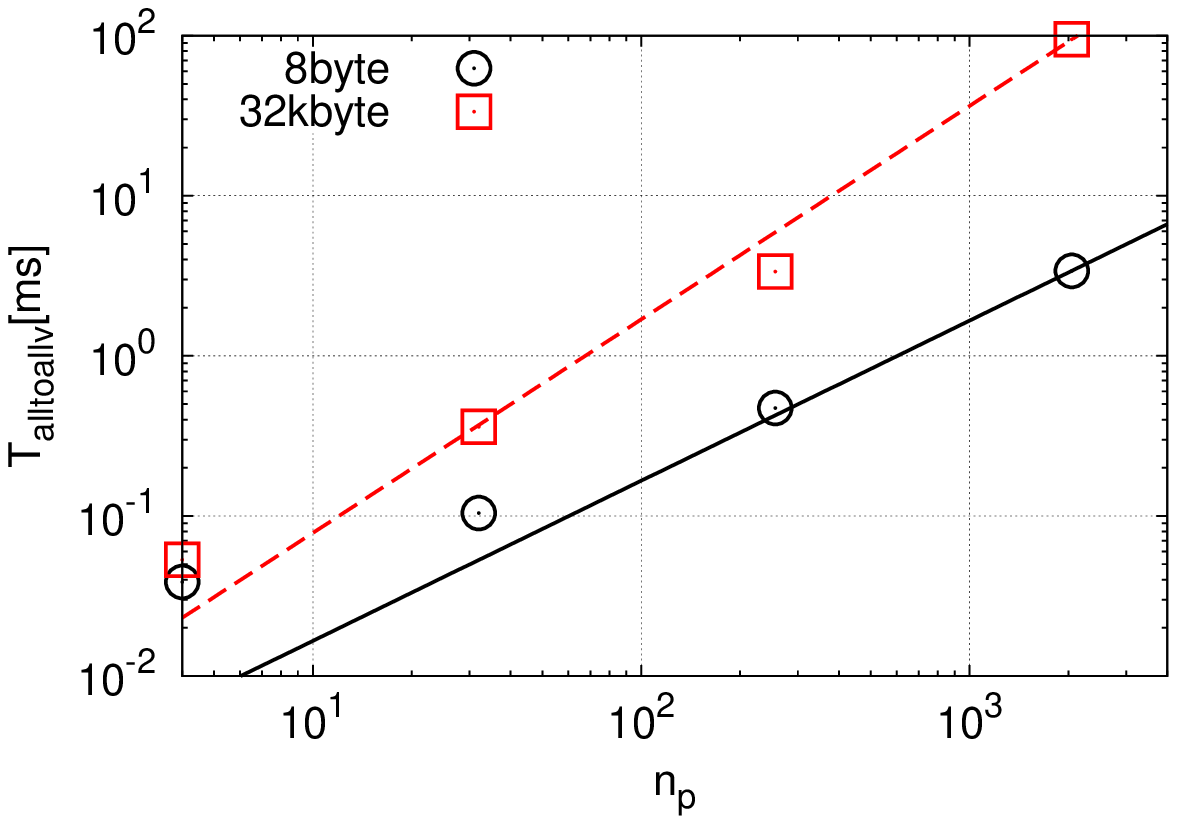}
  \end{center}
  \caption{
  
    Elapsed time of {\tt MPI\_Alltoallv} to send messeage of 8 bytes
    (circles) and 32k bytes (squares) as a function of the number of
    processes measured on K computer. Solid and dashed curves indicate
    the results for the message size of 8 bytes and 32k bytes,
    respectively.
    
} \label{fig:group_comm_a2a_2}
\end{figure}

The coefficients themselves in equation (\ref{eq:wtime_group_comm})
depend on the number of MPI processes $n_p$, as shown in
figures \ref{fig:group_comm_a2a_2}. They are modeled as
\begin{eqnarray}
\label{eq:alltoallv}
  T_{\rm alltoallv,startup} &=& \tau_{\rm alltoallv,startup} n_p, \\
  T_{\rm alltoallv,word} &=& \tau_{\rm alltoallv,word} n_p^{4/3}.
\end{eqnarray}
Here we assume that the speed to transfer message using {\tt
MPI\_Alltoallv} is limited to the bisection bandwidth of the system.
Under this assumption, $T_{\rm alltoallv,word}$ should be proportional
to $n_p^{4/3}$.  To estimate $\tau_{\rm alltoallv,startup}$ and
$\tau_{\rm alltoallv,word}$, we use measurements for message sizes of
8 bytes and 32k bytes. In K computer, we found that $\tau_{\rm
alltoallv,startup}$ is $0.00166$ ms and $\tau_{\rm alltoallv,word}$ is
$1.11 \times 10^{-7}$ ms per byte. If {\tt MPI\_Alltoallv} is limited
to the bisection bandwidth in K computer, $\tau_{\rm alltoallv,word}$
would be $5 \times 10^{-8}$ ms per byte. We can see that the actual
performance of {\tt MPI\_Alltoallv} on K computer is quite good.

\subsection{Domain decomposition}

For the hierarchical domain decomposition method described in
section \ref{sec:decomposition}, the calculation time is expressed as

\begin{equation}
  \label{eq:dccost}
  T_{\rm dc} = T_{\rm dc,gather}
            +   T_{\rm dc,sort}
            +   T_{\rm dc,exch}
            +   T_{\rm dc,misc},
\end{equation}
where $T_{\rm dc,gather}$ is the time for the $(i,0,0)$ process to
collect sample particles, $T_{\rm dc,sort}$ is the time to sort sample
particles on the $(i,0,0)$ process, $T_{\rm dc,exch}$ is the time to
exchange particles after the new domains are determined, and $T_{\rm
dc,misc}$ is the time for remaining procedures such as initial
exchange of samples in $x$ direction, exchange of sample particles and
domain boundaries in $x$ direction, and broadcasting of the domain
boundaries in $y$-$z$ planes.

On the machines we so far tested, $T_{\rm dc,gather}$ and $T_{\rm
  dc,misc}$ are much smaller than $T_{\rm dc,sort}$ and $T_{\rm
  dc,exch}$. Therefore we consider these two terms only.

First, we consider the time to sort sample particles. Since we use the
quick sort, the term $T_{\rm dc,sort}$ is expressed as
\begin{eqnarray}
T_{\rm dc,sort} &=& \tau_{\rm sort} \left[ 2n_{\rm smp} n_y n_z {\rm log} (n_{\rm smp} n_y n_z) + n_y n_{\rm smp} n_z {\rm log} (n_{\rm smp} n_z) \right] \\ 
             &\sim& \tau_{\rm dc,sort} n_{\rm smp}n_p^{2/3},
\end{eqnarray}
where $n_{\rm smp}$ is the average number of sample particles per
process, and $n_x$, $n_y$ and $n_z$ are the numbers of processes in x,
$y$ and $z$ direction. Here, $\tau_{\rm dc,sort} \sim {\rm log}(n_{\rm
smp}^3n_p^{5/3})\tau_{\rm sort}$. The first term expresses the time to
sort samples in $y$-$z$ planes with respect to $x$ and $y$
directions. The second term expresses that time to sort samples
respect to $z$ direction.

In order to model $T_{\rm dc,exch}$, we need to model the number of
particles which moves from one domain to another. This number would
depend on various factors, in particular the nature of the system we
consider. For example, if we are calculating the early phase of the
cosmological structure formation, particles do not move much in a
single timestep, and thus the number of particles moved between
domains is small. On the other hand, if we are calculating single
virialized self-gravitating system, particles move a relatively large
distances (comparable to average interparticle distance) in a single
timestep. In this case, if one process contains $n$ particles, half of
particles in the ``surface'' of the domain might migrate in and out
the domain. Thus, $O(n^{2/3})$ particles could be exchanged in this
case.

\begin{figure}
  \begin{center}
    \includegraphics[width=8cm]{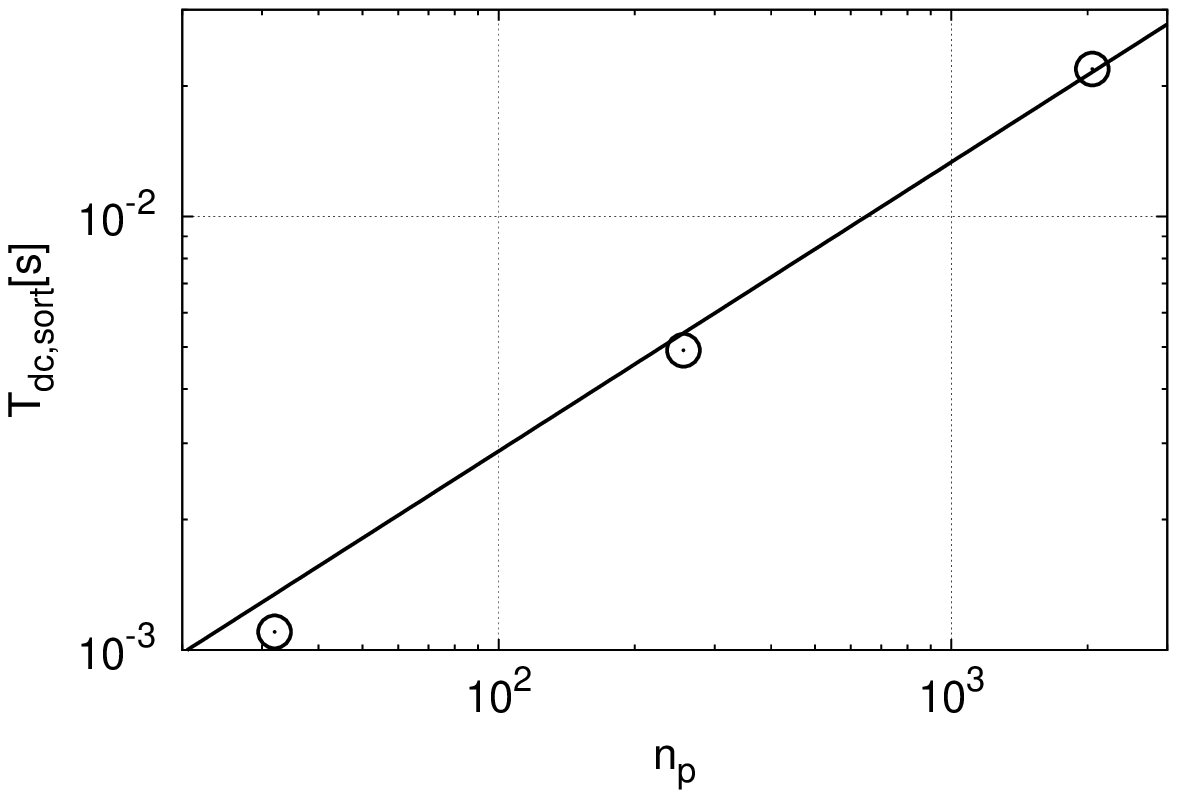}
  \end{center}
  \caption{

Measured $T_{\rm dc,sort}$ and its analytic model as a function of
$n_p$, in the case of $n_{\rm smp}=500$ and $n \sim 5.3 \times 10^5$.

    }
  \label{fig:domain_decomposition_sort}
\end{figure}

\begin{figure}
  \begin{center}
    \includegraphics[width=8cm]{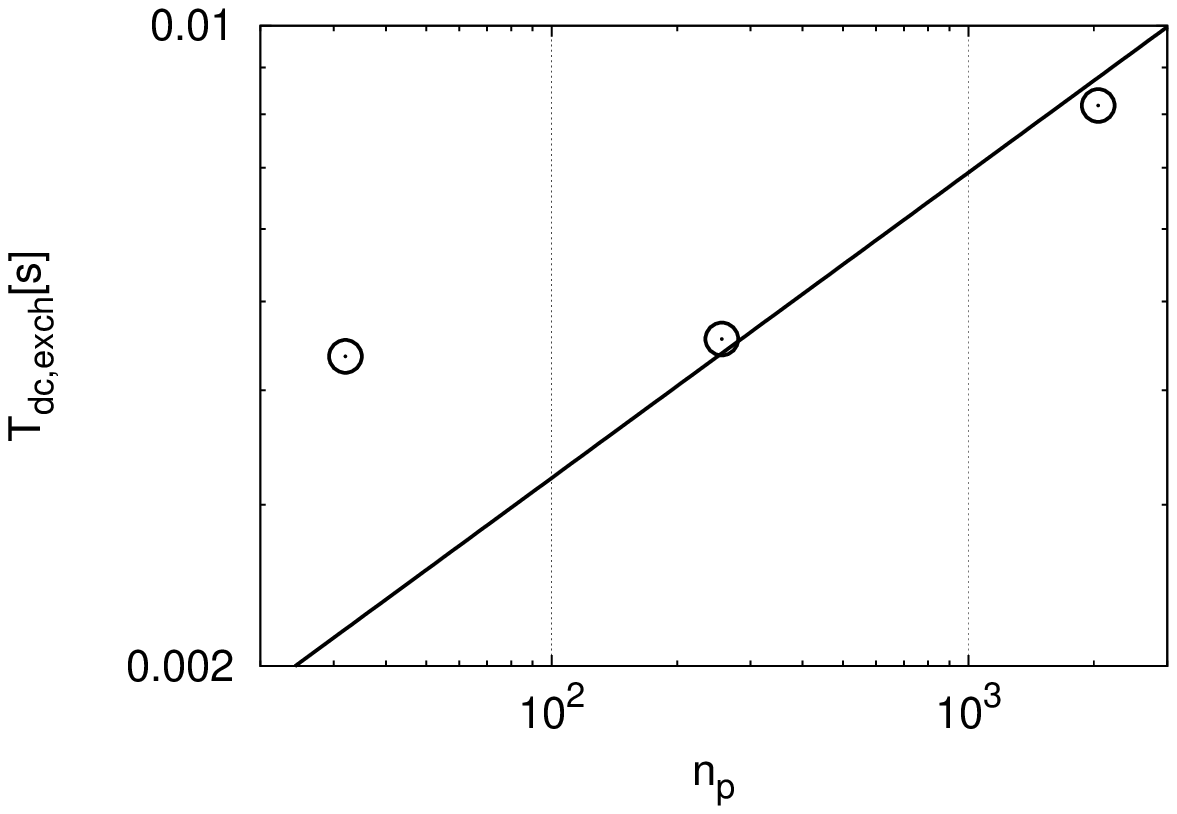}
  \end{center}
  \caption{

Measured $T_{\rm dc,exch}$ and its analytic model as a function of $n_p$, in
the case of $n_{\rm smp}=500$ and $n \sim 5.3 \times 10^5$.

    }
  \label{fig:domain_decomposition_exch}
\end{figure}

\begin{figure}
  \begin{center}
    \includegraphics[width=8cm]{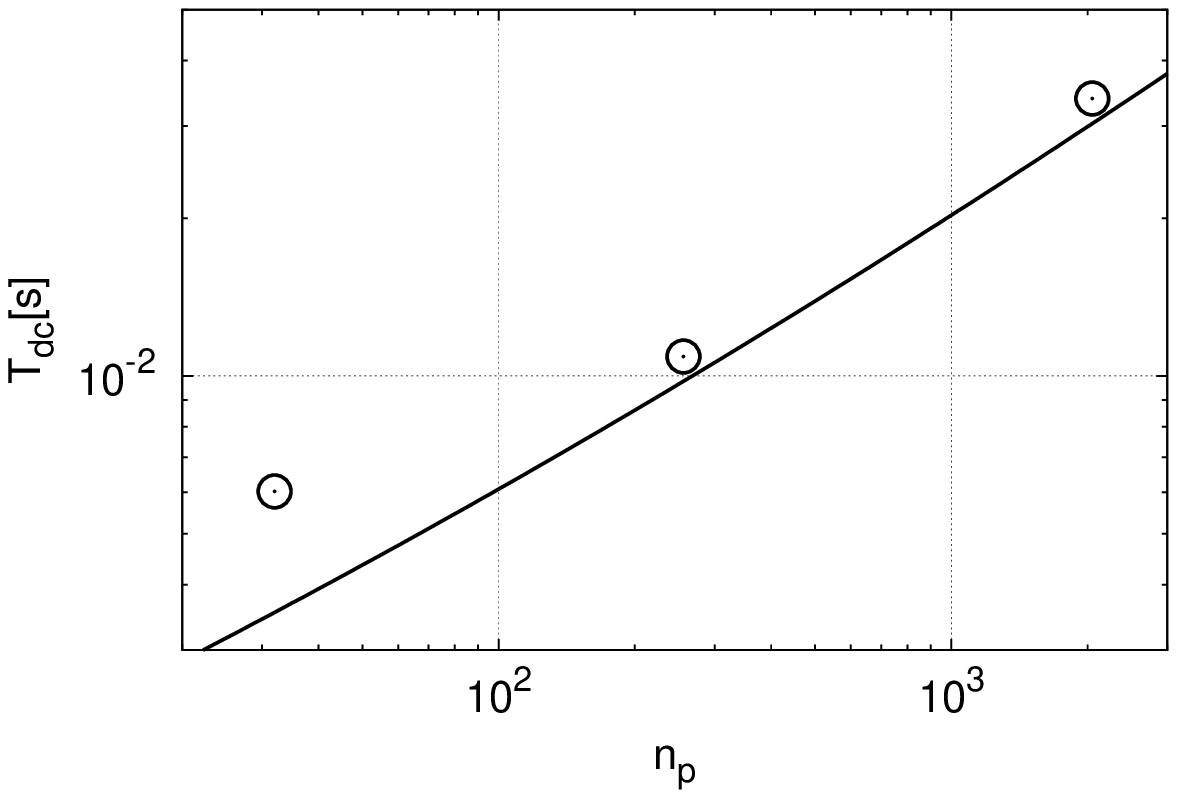}
  \end{center}
  \caption{

Measured $T_{\rm dc}$ and its analytic model as a function of $n_p$,
in the case of $n_{\rm smp}=500$ and $n \sim 5.3 \times 10^5$.

    }
  \label{fig:domain_decomposition}
\end{figure}

Figures \ref{fig:domain_decomposition_sort}, \ref{fig:domain_decomposition_exch}
and \ref{fig:domain_decomposition} show the elapsed time for sorting
samples, exchanging samples, and domain decomposition for the case of
disk galaxy simulations in the case of $n_{\rm smp}=500$ and $n \sim
5.3 \times 10^5$. We also plot the analytic models given by
\begin{eqnarray}
\label{eq:ddfit}
  T_{\rm dc} &\sim& T_{\rm dc,sort} + T_{\rm dc,exch} \\
             &=& \tau_{\rm dc,sort} n_{\rm smp} n_p^{2/3}
                 + \tau_{\rm dc,exch} \sigma \Delta t / \left<r\right> n^{2/3} b_p,
\end{eqnarray}
where $\tau_{\rm dc,sort}$ and $\tau_{\rm dc,exch}$ are the execution
time for sorting one particle and for exchanging one particle
respectively, $\sigma$ is the typical velocity of particles, $\Delta
t$ is the timestep and $\left<r\right>$ is the average interparticle
distance. For simplicity we ignore weak log term in $T_{\rm dc,sort}$
. On K computer, $\tau_{\rm dc,sort} = 2.67\times 10^{-7}$ second and
$\tau_{\rm dc,exch} = 1.42\times 10^{-7}$ second per byte. Note that
$\tau_{\rm dc,exch} \sim 672 T_{\rm p2p,word}$.

In figure \ref{fig:domain_decomposition_exch}, for small $n_p$, the
analytic model gives the value about 2 times smaller than the measured
point. This is because the measured values include not only the time
to exchange particles but also the time to determine appropriate
processes to send particles, while the analytic model includes only
the time to exchange particles. For small $n_p$, the time to determine
the appropriate process is not negligible, and therefor the analytic
model gives an underestimate.

The analysis above, however, indicates that $T_{\rm dc,exch}$ is, even
when it is relatively large, still much smaller than $T_{\rm exch}$,
which is the time to exchange particles and superparticles for
interaction calculation (see section \ref{sec:exchnage_list}).

\subsection{Tree construction}

Theoretically, the cost of tree construction is $O(n{\rm log}n)$, and
of the same order as the interaction calculations itself. However, in
our current implementation, the interaction calculation is much more
expensive, independent of target architecture and the type of the
interaction. Thus we ignore the time for the tree constructions.

\subsection{Exchange of particles and superparticles}
\label{sec:exchnage_list}

For the exchange of particles and superparticles, in the current
implementation of FDPS, first each node constructs the list of
particles and superparticles (hereafter the exchange list) to be sent
to all other nodes, and then data are exchanged through a single call
to {\tt MPI\_Alltoallv}. The way the exchange list is constructed
depends on the force calculation mode. In the case of long-range
forces, usual tree traversal with a fixed opening angle $\theta$ is
performed. For the short-range forces, the procedure used depends on
the subtypes of the interaction. In the case of fixed or $j$-dependent
cutoff, the exchange list for a node can be constructed by a single
traversal of the local tree. On the other hand, for $i$-dependent or
symmetric cutoff, first each node constructs the $j$-dependent
exchange lists and sends them to all other nodes. Each node then
constructs the $i$-dependent exchange lists and sends them again.

The time for the construction and exchange of exchange list is thus
given by
\begin{equation}
  T_{\rm exch} = k_{\rm type}(T_{\rm exch,const}+T_{\rm exch,comm}).
  \label{eq:exchangecost}
\end{equation}

Here, $k_{\rm type}$ is an coefficient which is unity for fixed and
$j$-dependent cutoffs and two for other cutoffs. Strictly speaking,
the communication cost does not double for $i$-dependent or symmetric
cutoffs, since we send only particles which were not sent in the first
step. However, for simplicity we use $k=2$ for both calculation and
communication.

\begin{figure}
  \begin{center}
    \includegraphics[width=8cm]{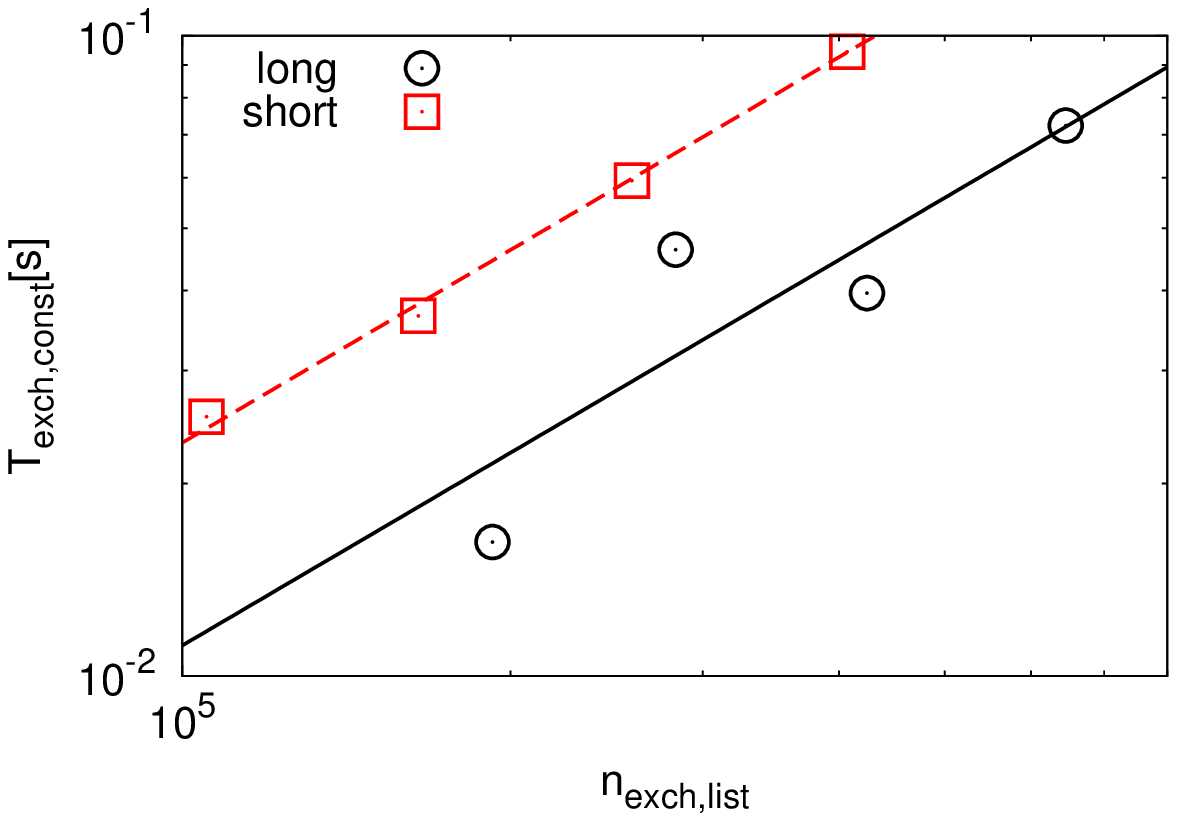}
  \end{center}
  \caption{
  
Time for the construction of the exchange list plotted against the
average length of the list, for the case of $n_p=2048$ and $n \sim
2.7 \times 10^5, 5.3 \times 10^5, 1.1 \times 10^6, 2.1 \times
10^6$. Circles and squares indicate the results for long-range and
short-range force, respectively. Solid and dashed curves are analytic
models [equation (\ref{eq:exchange})].

    }
  \label{fig:exchangeLET_nlist-wtime_const}
\end{figure}

\begin{figure}
  \begin{center}
    \includegraphics[width=8cm]{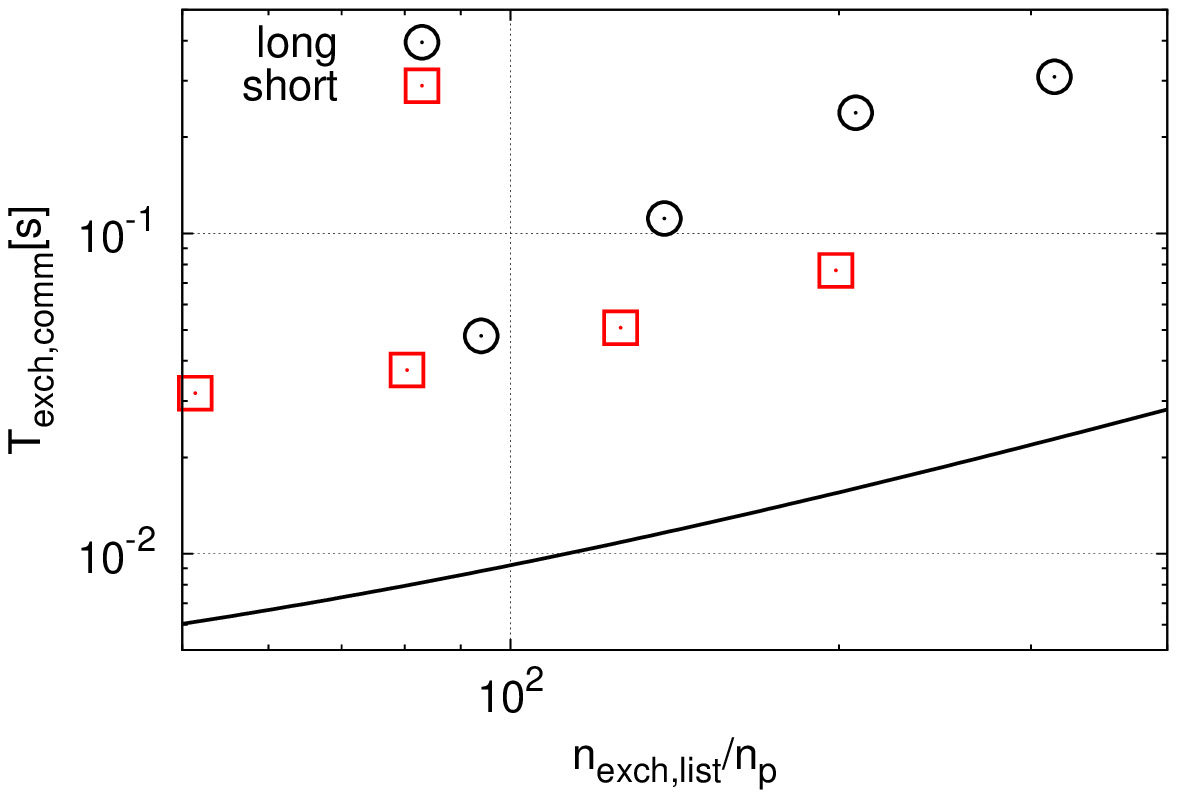}
  \end{center}
  \caption{

Time for the communication of the exchange list against the average
length of the list per process, for the case of $n_p=2048$ and $n \sim
2.7 \times 10^5, 5.3 \times 10^5, 1.1 \times 10^6, 2.1 \times
10^6$. Circles and squares indicate the results for long-range and
short-range force, respectively. The curve is predicted from
equation (\ref{eq:exchange}).

    }
  \label{fig:exchangeLET_nlist-wtime_exch}
\end{figure}

The two terms in equation (\ref{eq:exchangecost}) are then
approximated as
\begin{eqnarray}
  T_{\rm exch,const} &=& \tau_{\rm exch,const} n_{\rm exch,list}, \label{eq:exchange} \\
  T_{\rm exch,comm}(n_{\rm msg}) &=&  T_{\rm alltoallv}\left( n_{\rm exch,list}/n_{p} b_p \right), \label{eq:exchange_comm}
\end{eqnarray}

where $n_{\rm exch,list}$ is the average length of the exchange list
and $\tau_{\rm exch,const}$ is the execution time for constructing one
exchange list.  Figures \ref{fig:exchangeLET_nlist-wtime_const}
and \ref{fig:exchangeLET_nlist-wtime_exch} show the execution time for
constructing and exchanging the exchange list against the average
length of the list. Here, $b_p=48$ bytes for both short and long-range
interactions. From figure \ref{fig:exchangeLET_nlist-wtime_const}, we
can see that the elapsed time can be fitted well by equation
(\ref{eq:exchange}).  Here $\tau_{\rm exch,const}$ is $1.12 \times
10^{-7}$ second for long-range interaction and $2.31 \times 10^{-7}$
second for short-range interaction.

From figure \ref{fig:exchangeLET_nlist-wtime_exch}, we can see a large
discrepancy between measured points and the curves predicted from
equation (\ref{eq:exchange_comm}). In the measurement of the
performance of ${\tt MPI\_Alltoallv}$ in section \ref{sec:comm_model},
we used uniform message length across all processes. In actual use in
exchange particles, the length of the message is not
uniform. Neighboring processes generally exchange large messages,
while distant processes exchange short message. For such cases,
theoretically, communication speed measured in terms of average
message length should be faster. In practice, however, we observed a
serious degradation of performance. This degradation seems to imply
that the current implementation of ${\tt MPI\_Alltoallv}$ is
suboptimal for non-uniform message size.

In the following, we estimate $n_{\rm exch,list}$. If we consider a
rather idealized case, in which all domains are cubes containing $n$
particles, the total length of the exchange lists for one domain can
approximately be given by
\begin{equation}
\label{eq:n_list_long}
  n_{\rm exch,list} \sim \frac{14n^{2/3}}{\theta} + \frac{21\pi n^{1/3}}{\theta^2} + \frac{28\pi}{3\theta^3} {\rm log_2} \left\{ \frac{\theta}{2.8}\left[\left(nn_p\right)^{1/3} - n^{1/3}\right] \right\},
\end{equation}
for the case of long-range interactions and
\begin{equation}
\label{eq:n_list_short}
  n_{\rm exch,list} \sim \left( n^{1/3}-1+2\frac{r_{\rm cut}}{\left<r\right>} \right)^3-n,
\end{equation}
for the case of short-range interactions, where $r_{\rm cut}$ is the
average cutoff length and and $\left<r\right>$ is the average
interparticle distance. In this section we set $r_{\rm cut}$ so that
the number of particles in the neighbor sphere is to be 100. In other
words, $r_{\rm cut} \sim 3\left<r\right>$.

% and $n_{\rm ngb}$ is the number of the
%particles in the sphere with a radius of $r_{\rm cut}$.

\begin{figure}
  \begin{center}
    \includegraphics[width=8cm]{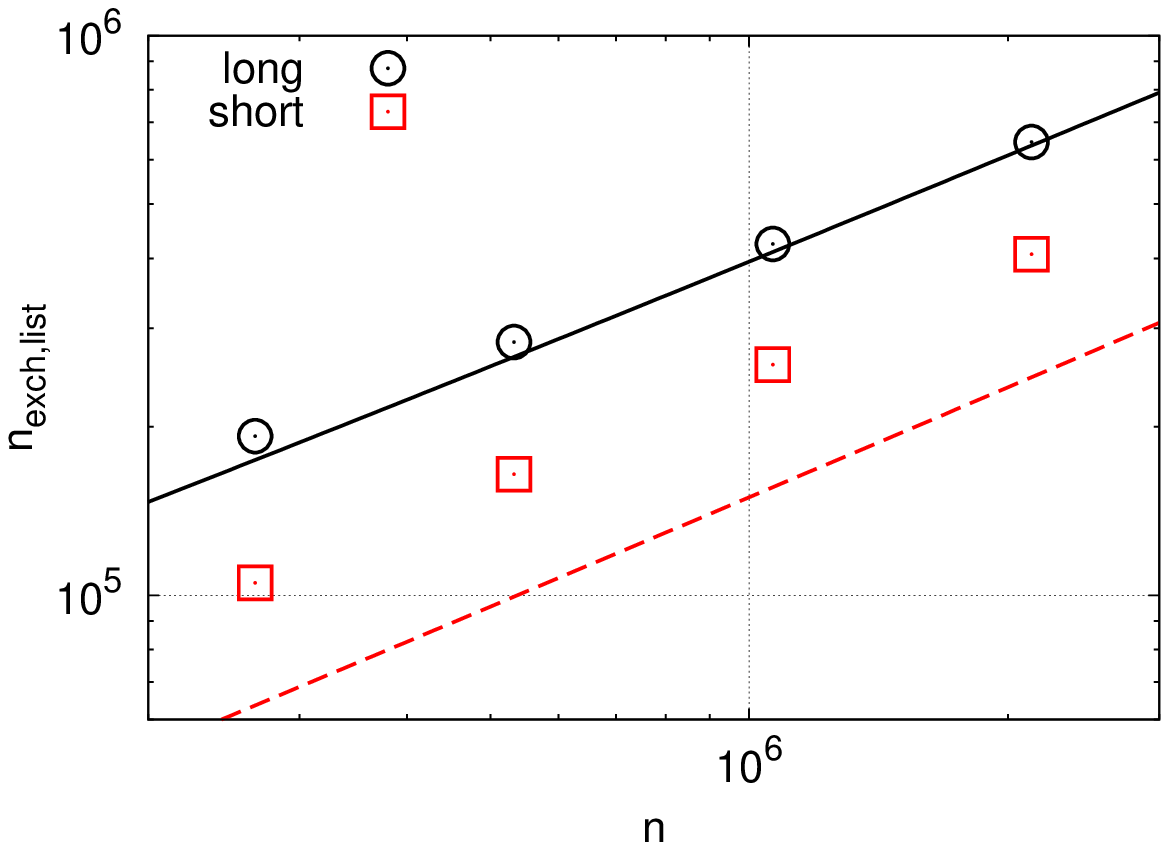}
  \end{center}
  \caption{
  
The average length of the exchange lists for long-range interaction
(circles) and for short-range interaction (squares) as a function of
$n$, in the case of $\theta=0.4$ and $n_P=2048$. Solid and dashed
curves are predicted from equations (\ref{eq:n_list_long}) and
(\ref{eq:n_list_short}), respectively.

    }
  \label{fig:exchangeLET_nlist-nsend}
\end{figure}

In figure \ref{fig:exchangeLET_nlist-nsend}, we plot the list length
for short and long interactions against the average number of
particles. The rough estimate of equations (\ref{eq:n_list_long}) and
(\ref{eq:n_list_short}) agree very well with the measurements.

%Two curves are from
%equation \ref{eq:n_list_long} and \ref{eq:n_list_short}.

\subsection{Tree traverse and interaction calculation}

The time for the force calculation is given by
\begin{equation}
T_{\rm icalc} = T_{\rm icalc,force} + T_{\rm icalc,const},
\end{equation}
where $T_{\rm icalc,force}$ and $T_{\rm icalc,const}$ are the time for
the force calculations for all particles and the tree traverses for
all interaction lists, respectively.

\begin{figure}
  \begin{center}
    \includegraphics[width=8cm]{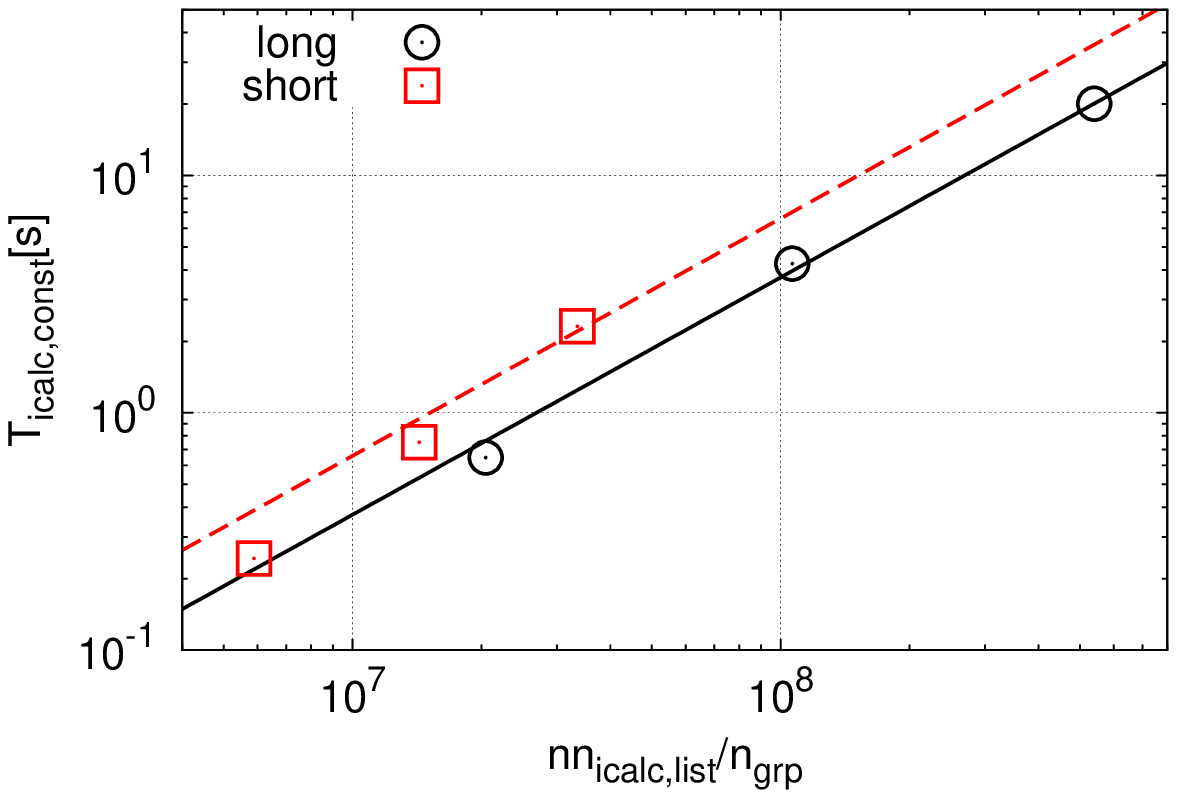}
  \end{center}
  \caption{

Time for the construction of the interaction list for long-range force
(circles) and short-range force (squares), for the case of $n \sim
5.3 \times 10^5$ and $\theta = 0.4$. Solid and dashed curves are the
analytic models for long-rage and short range forces, respectively
[equation (\ref{eq:wtime_force})].

    }
  \label{fig:interaction_calc_nwalk-wtime}
\end{figure}

$T_{\rm icalc,force}$ and $T_{\rm icalc,const}$ are expressed as
\begin{eqnarray}
T_{\rm icalc,const} &=& \tau_{\rm icalc,const} n n_{\rm icalc,list} / n_{\rm grp}, \label{eq:wtime_force} \\
T_{\rm icalc,force}  &=& \tau_{\rm icalc,force} n n_{\rm icalc,list},
\end{eqnarray}
where $n_{\rm icalc,list}$ is the average length of the interaction
list, $n_{\rm grp}$ is the number of $i$ particle groups for modified
tree algorithms by \citet{1990JCoPh..87..161B}, $\tau_{\rm
icalc,force}$ and $\tau_{\rm icalc,const}$ are the time for one force
calculation and for constructing one interaction list.  In
figure \ref{fig:interaction_calc_nwalk-wtime}, we plot the time for
the construction of the interaction list as a function of $n_{\rm
grp}$. On K computer, $\tau_{\rm icalc,const}$ are $3.72\times
10^{-8}$ second for the long-range force and $6.59\times 10^{-8}$
second for the short-range force.

The length of the interaction list is given by
\begin{equation}
\label{eq:nlist_long}
  n_{\rm icalc,list} \sim n_{\rm grp} + \frac{14n_{\rm grp}^{2/3}}{\theta} + \frac{21\pi n_{\rm grp}^{1/3}}{\theta^2} + \frac{28\pi}{3\theta^3} {\rm log_2} \left[ \frac{\theta}{2.8}\left\{\left(nn_p\right)^{1/3} - n_{\rm grp}^{1/3}\right\}\right]
\end{equation}
for the case of long-range interactions and
\begin{equation}
\label{eq:nlist_short}
  n_{\rm icalc,list} \sim \left( n_{\rm grp}^{1/3}-1+2\frac{r_{\rm cut}}{\left<r\right>} \right)^3,
\end{equation}
for the case of short-range interactions.

\begin{figure}
  \begin{center}
    \includegraphics[width=8cm]{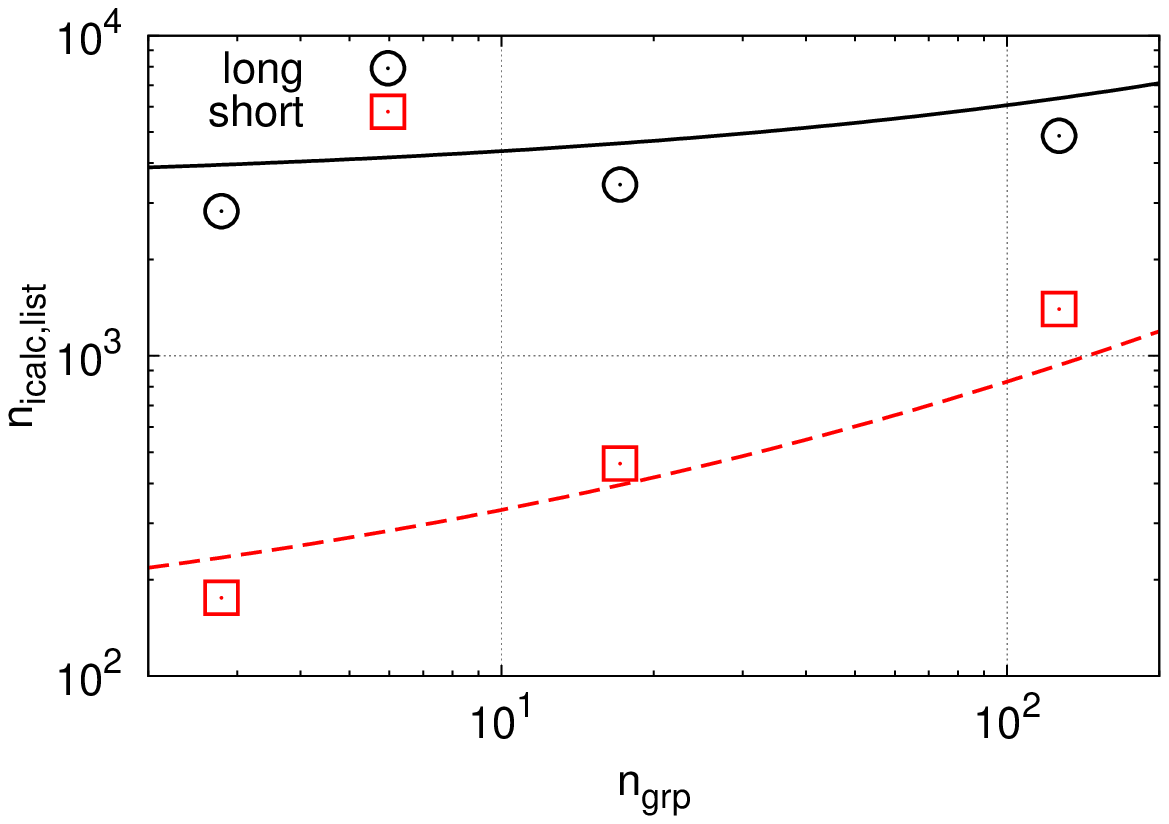}
  \end{center}
  \caption{

The average length of the interaction list for long-range force
(circles) and short-range force (squares), for the case of $n \sim
5.3 \times 10^5$ and $\theta = 0.4$. Solid and dashed curves are
analytic models for long-range [equation (\ref{eq:nlist_long})] and
short-range [equation (\ref{eq:nlist_short})] forces.

    }
  \label{fig:interaction_calc_ng-nlist}
\end{figure}

In figure \ref{fig:interaction_calc_ng-nlist}, we plot the length of
the interactions lists for long-range force and short-range force. We
can see that the length of the interaction lists can be fitted
reasonably well.

\begin{figure}
  \begin{center}
    \includegraphics[width=8cm]{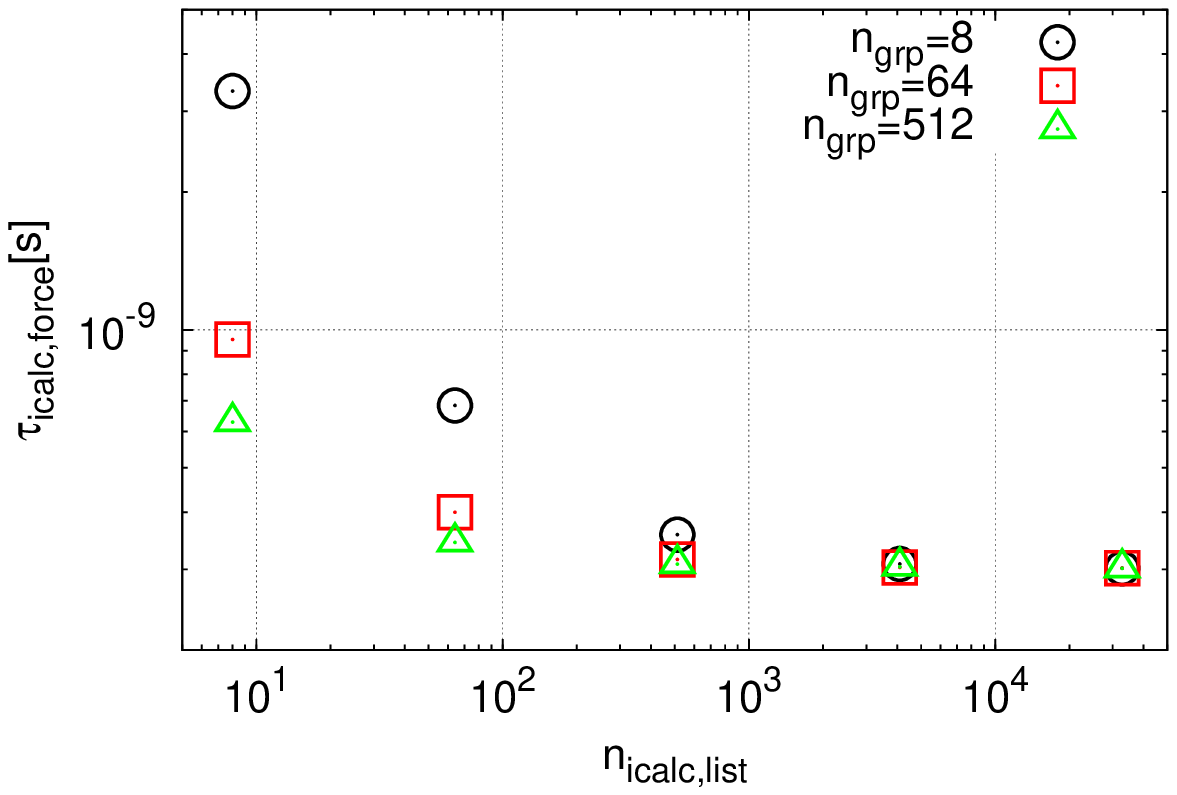}
  \end{center}
  \caption{
  
  Time for the evaluation of one gravity force against $n_{\rm icalc,
  list}$ for various $n_{\rm grp}$.
  
    }
  \label{fig:calc_force}
\end{figure}

In the following, we discuss the time for the force calculation.  The
time for the force calculation for one particle pair $\tau_{\rm
icalc,force}$ has different values for different kinds of
interactions.  We plot $\tau_{\rm icalc,force}$ against $n_{\rm
icalc,list}$ for various $n_{\rm grp}$ in
figure \ref{fig:calc_force}. We can see that for larger $n_{\rm grp}$,
$\tau_{\rm icalc,force}$ becomes smaller. However, from equation
(\ref{eq:nlist_long}), large $n_{\rm grp}$ leads to large $n_{\rm
icalc,list}$ and the number of interactions becomes larger. Thus there
is an optimal $n_{\rm grp}$. In our disk-galaxy simulations in K
computer, the optimal $n_{\rm grp}$ is a few hundreds, and dependence
on $n_{\rm grp}$ is weak.

\subsection{Total time}
\label{sec:total_time}

Now we can predict the total time of the calculation using the above
discussions. The total time per one timestep is given by
\begin{eqnarray}
  T_{\rm step} &\sim& T_{\rm dc,sort}/n_{\rm dc} \nonumber
                  + k_{\rm type}\left( T_{\rm exch,const}+T_{\rm exch,comm} \right) \nonumber \\
                  &&+ T_{\rm icalc,force} + T_{\rm icalc,const} \label{eq:totalcost2} \\
             &\sim& \tau_{\rm dc,sort} n_{\rm smp} n_p^{2/3}/n_{\rm dc} \nonumber \\
                &&+ k_{\rm type}\left( \tau_{\rm exch,const} n_{\rm exch,list} + \tau_{\rm alltoallv,startup}n_p + \tau_{\rm alltoallv,word}n_{\rm exch,list}b_pn_p^{1/3} \right) \nonumber \\
                && + \tau_{\rm icalc,force} n n_{\rm icalc,list} \nonumber \\
                && + \tau_{\rm icalc,const} n n_{\rm icalc,list} / n_{\rm grp}.   \label{eq:totalcost3}
\end{eqnarray}
The time coefficients in equation (\ref{eq:totalcost3}) for K computer
are summarized in table \ref{table:time_coefficients}. In this section
we use $n_{\rm dc}=1$.

\begin{table}
\caption{Time coefficients in equation (\ref{eq:totalcost3}) for K computer.
$\tau_{\rm icalc,force}$ is the value for gravity.  }
\begin{tabular}{|l|l|} \hline
$\tau_{\rm alltoallv,startup}$ [s] & $1.66 \times 10^{-6}$ \\ 
$\tau_{\rm alltoallv,word}$ [s/byte] & $1.11 \times 10^{-10}$\\
$\tau_{\rm dc,sort}$ [s] & $2.67\times 10^{-7}$ \\ 
$\tau_{\rm exch,const}$ [s] & $1.12 \times 10^{-7}$ \\
$\tau_{\rm icalc,const}$ [s] & $3.72\times 10^{-8}$ \\
$\tau_{\rm icalc,force}$ [s] & $3.05\times 10^{-10}$ \\ \hline
\end{tabular}
\label{table:time_coefficients}
\end{table}

\begin{figure}
  \begin{center}
    \includegraphics[width=8cm]{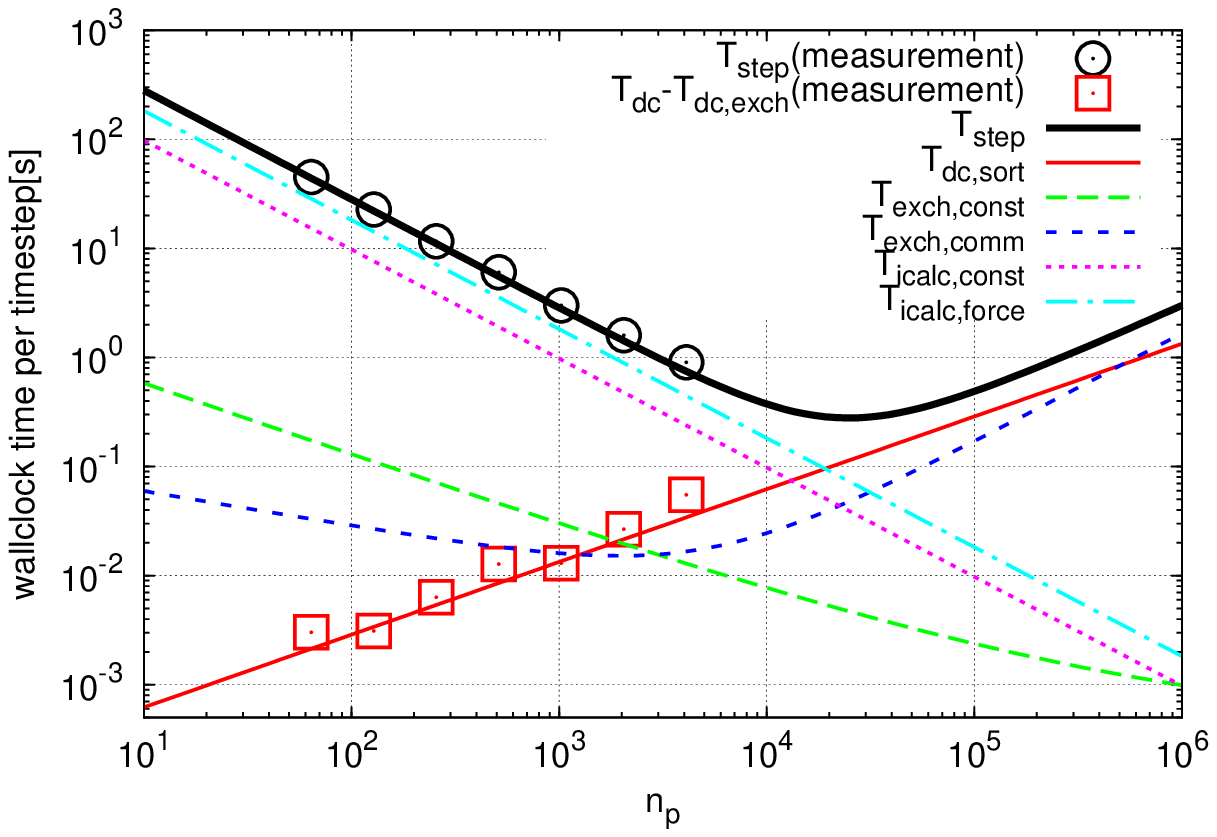}
  \end{center}
  \caption{
  
  Breakdown of the total time of the calculation per one timestep
  against $n_p$, for the case of $N=5.5\times 10^8$ , $n_{\rm
  smp}=500$, $\theta=0.4$ and $n_{\rm grp}=130$.
  
}
  \label{fig:performance_model_strong_breakdown_comp}
\end{figure}

To see if the predicted time by equation (\ref{eq:totalcost3}) is
reasonable, we compare the predicted time and the time obtained from
the disk galaxy simulation with the total number of particles ($N$) of
550 million and $\theta = 0.4$. In our simulations, we use up to the
quadrupole moment. On the other hand, we assume the monopole moment
only in equation (\ref{eq:totalcost3}). Thus we have to correct the
time for the force calculation in equation (\ref{eq:totalcost3}). In
our simulations, the cost of the force calculation of the quadrupole
moment is two times higher than that of the monopole moment and about
75\% of particles in the interactions list are superparticles.  Thus
the cost of the force calculation in the simulation is 75\% higher
than the prediction. We apply this correction to equation
(\ref{eq:totalcost3}). In
figure \ref{fig:performance_model_strong_breakdown_comp}, we plot the
breakdown of the predicted time with the correction and the obtained
time from the disk galaxy simulations. We can see that our predicted
times agree with the measurements very well.

\begin{figure}
  \begin{center}
    \includegraphics[width=8cm]{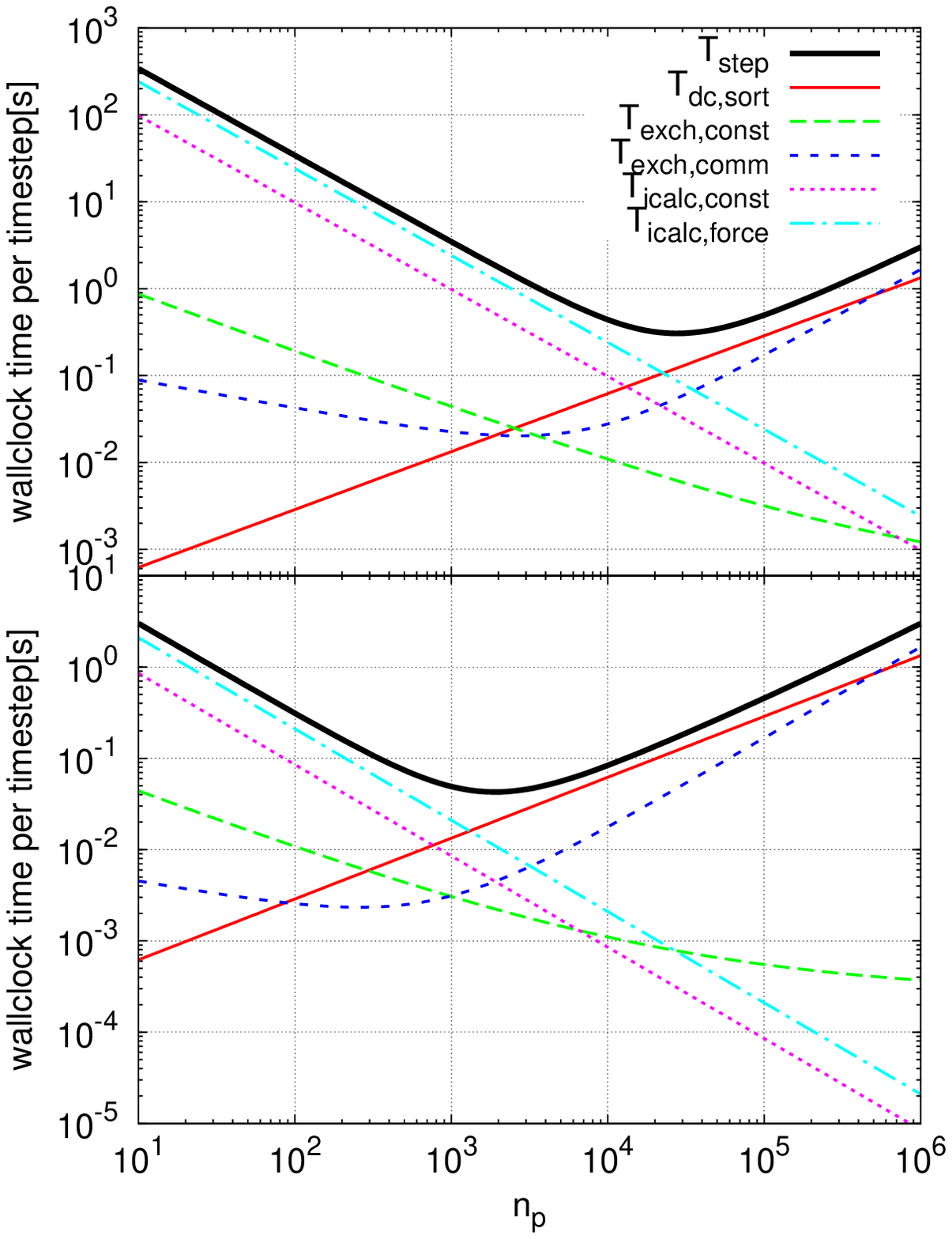}
  \end{center}
  \caption{
  
  Breakdown of the total time of the calculation per one timestep
  against $n_p$, for the case of $N=10^9$ (top panel) and $=10^7$
  (bottom panel), $n_{\rm smp}=500$, $\theta=0.4$ and $n_{\rm
  grp}=300$.
  
}
  \label{fig:performance_model_strong_breakdown}
\end{figure}

In the following, we analyze the performance of the gravitational many
body simulations for various hypothetical computers. In
figure \ref{fig:performance_model_strong_breakdown}, we plot the
breakdown of the calculation time predicted using equation
(\ref{eq:totalcost3}) for the cases of 1 billion and 10 million
particles against $n_p$. For the case of 1 billion particles, we can
see that the slope of $T_{\rm step}$ becomes shallower for
$n_p \gtrsim 10000$ and increases for $n_p \gtrsim 30000$, because
$T_{\rm dc,sort}$ dominates. Note that $T_{\rm exch,comm}$ also has
the minimum value. The reason is as follows. For small $n_p$, $T_{\rm
alltoallv,word}$ is dominant in $T_{\rm exch,comm}$ and it decrease as
$n_p$ increases, because the length of $n_{\rm exch,list}$ becomes
smaller. For large $n_p$, $T_{\rm alltoallv,startup}$ becomes dominant
and it increases linearly. We can see the same tendency for the case
of 10 million particles. However, the optimal $n_p$, at which $T_{\rm
step}$ is the minimum, is smaller than that for 1 billion particles,
because $T_{\rm dc,sort}$ is independent of $N$.

\begin{figure}
  \begin{center} \includegraphics[width=8cm]{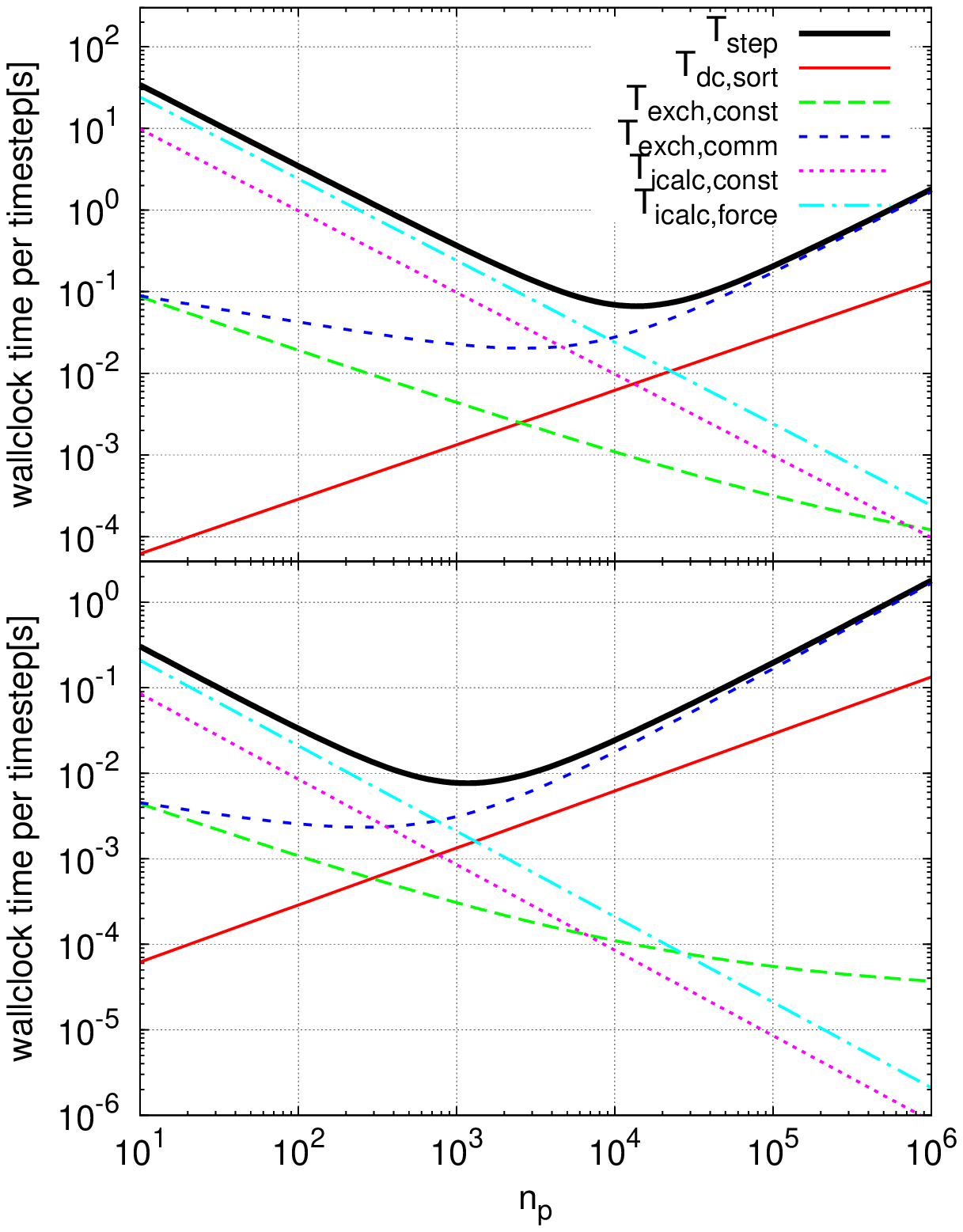} \end{center} \caption{
  
The same as figure \ref{fig:performance_model_strong_breakdown}, but
for the floating-point operation performance ten times faster than K
computer.
  
}
  \label{fig:performance_model_strong_breakdown_x10}
\end{figure}

In figure \ref{fig:performance_model_strong_breakdown_x10}, we plot
the breakdown of the predicted calculation time for a hypothetical
computer which has the floating-point operation performance ten times
faster than that of K computer (hereafter X10). In other words,
$\tau_{\rm alltoallv,startup}$ and $\tau_{\rm alltoallv,word}$ are the
same as those of K computer, but $\tau_{\rm dc,sort}$, $\tau_{\rm
exch,const}$, $\tau_{\rm icalc,const}$ and $\tau_{\rm icalc,force}$
are ten times smaller than those of K computer. We can see that the
optimal $n_p$ is shifted to smaller $n_p$ for both cases of $N$ of 1
billion and 10 million, because $T_{\rm exch,comm}$ is unchanged.
However, the shortest time per timestep is improved by about a factor
of five.  If the network performance is also improved by a factor of
ten, we would get the same factor of ten improvement for the shortest
time per timestep. In other words, by reducing the network performance
by a factor of ten, we suffer only a factor of two degradation of the
shortest time.

\begin{figure}
  \begin{center}
    \includegraphics[width=8cm]{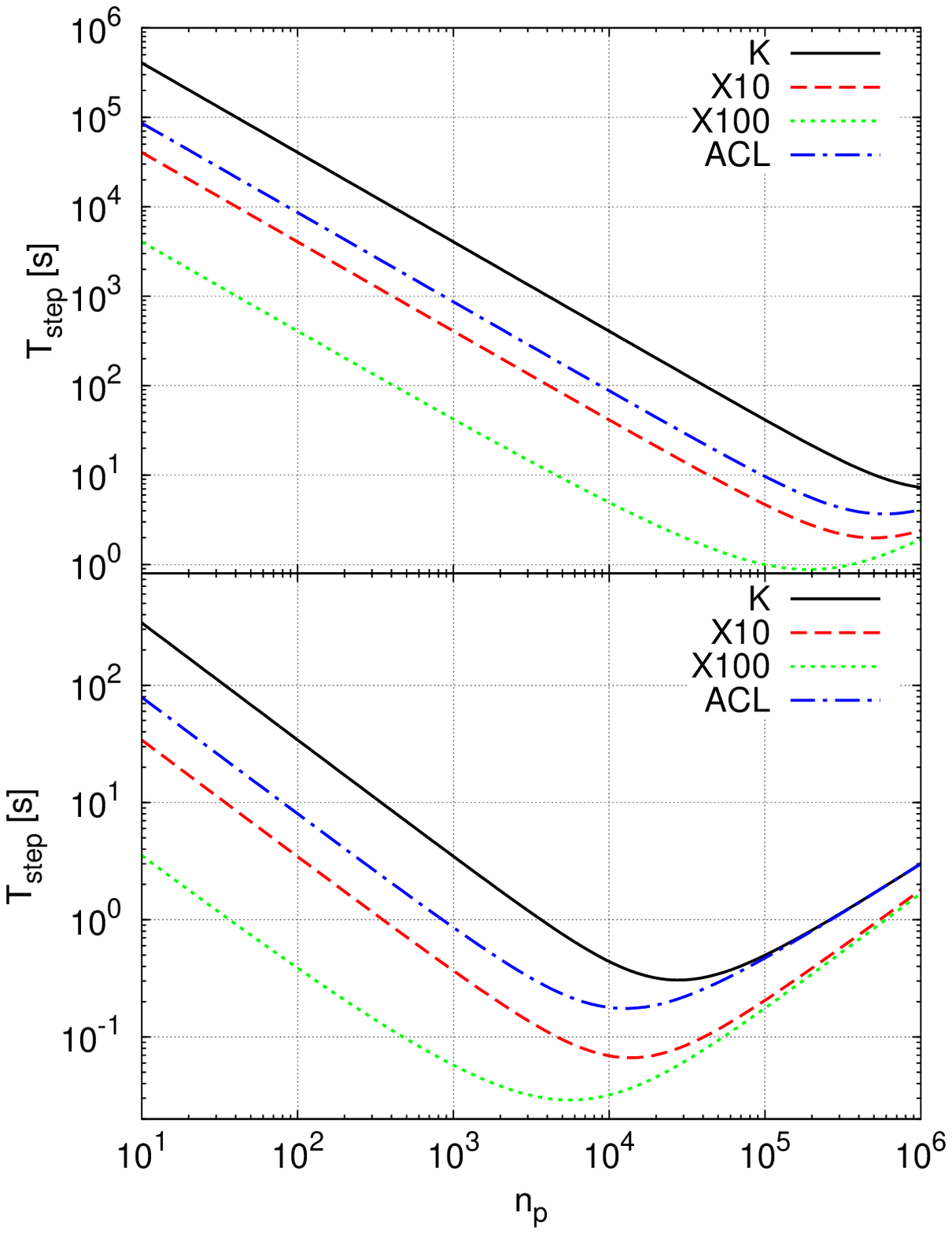}
  \end{center}
  \caption{
  
Predicted total calculation time for three hypothetical computers and
K computer as a function of $n_p$, for the case of $n_{\rm smp}=500$,
$\theta=0.4$.  Top and bottom panels indicate the results of the case
for $N=10^{12}$ and $N=10^{9}$, respectively.
  
}
  \label{fig:performance_model_strong_predict}
\end{figure}

In figure \ref{fig:performance_model_strong_predict}, we plot
predicted $T_{\rm step}$ for three hypothetical computers and K
computer. Two of four computers are the same computer models we used
above. Another is a computer with the floating-point operation
performance hundred times faster than K computer (hereafter X100).
The last one is a computer of which the performance of the force
calculation is ten times faster than K computer (hereafter ACL). In
other words, only $\tau_{\rm icalc,force}$ is ten times smaller than
that of K computer.  This computer is roughly mimicking a computer
with an accelerator such as, GPU \citep{hamada2009novel},
GRAPE \citep{1990Natur.345...33S, 2003PASJ...55.1163M} and
PEZY-SC. Here we use the optimal $n_{\rm grp}$, at which $T_{\rm
step}$ is minimum, for each computers. For the case of $N=10^9$, the
optimal $n_{\rm grp} \sim 300$ for K computer and X10, $\sim 400$ for
X100 and $\sim 1600$ for ACL. For the case of $N=10^{12}$, the optimal
$n_{\rm grp}$ for K, X10, X100 is the same as those for $N=10^9$, but
$\sim 1800$ for ACL. The optimal value of $n_{\rm grp}$ for ACL is
larger than those of any other computers, because large $n_{\rm grp}$
reduces the cost of the construction of the interaction list.

From figure \ref{fig:performance_model_strong_predict}, we can see
that for small $n_p$, X10 and X100 are ten and hundred times faster
than K computer, respectively. However, for the case of $N=10^9$,
$T_{\rm step}$ of the values of X10 and X100 increase for $n_p \gtrsim
15000$ and $\gtrsim 7000$, because the $T_{\rm exch,comm}$ becomes the
bottleneck. ACL shows a similar performance to that of X10 up to
optimal $n_p$, because the force calculation is dominant in the total
calculation time. On the other hand, for large $n_p$, the performance
of ACL is almost the same as that of K computer, because ACL has the
same bottleneck as K computer has, which is the communication of the
exchange list.  On the other hand, for the case of $N=10^{12}$,
$T_{\rm step}$ is scaled up to $n_p \sim 10^5$ for all computers. This
is because for larger $N$ simulations, the costs of the force
calculation and the construction of the interaction list are
relatively higher than the communication of the exchange list. Thus
the optimal $n_p$ is sifted to larger value if we use larger $N$.

From figures \ref{fig:performance_model_strong_breakdown}
and \ref{fig:performance_model_strong_breakdown_x10}, we can see that
for large $n_p$, performance will be limited by $T_{\rm dc,sort}$ and
$T_{\rm exch,comm}$. Therefor, if we can reduce them further, we can
improve the efficiency of the calculation with large $n_p$. It is
possible to reduce the time for sort by applying the algorithm used in
$x$ direction to $y$ direction as well or setting $n_{\rm dc}$ to more
than unity. It is more difficult to reduce $T_{\rm exch,comm}$, since
we are using system-provided {\tt MPI\_Alltoallv}.

\section{Conclusion}
\label{sec:conclusion}

In this paper, we present the basic idea, implementation, measured
performance and performance model of FDPS, a framework for developing
efficient parallel particle-based simulation codes.  FDPS provides all
of these necessary functions for the parallel execution of
particle-based simulations. By using FDPS, researchers can easily
develop their programs which run on large-scale parallel
supercomputers. For example, a simple gravitational $N$-body program
can be written in around 120 lines.

We implemented three astrophysical applications using FDPS and
measured their performances. All applications showed good performance
and scalability. In the case of the disk galaxy simulation, the
achieved efficiency is around 50\% of the theoretical peak, for the
cosmological simulation 7\%, and for the giant impact simulation 40\%.

We constructed the performance model of FDPS and analyzed the
performance of applications using FDPS. We found that the performance
for small number of particles would be limited by the time for the
calculation necessary for the domain decomposition and communication
necessary for the interaction calculation.

\bigskip

We thank M. Fujii for providing initial conditions of spiral
simulations, T. Ishiyama for providing his Particle Mesh code,
K. Yoshikawa for providing his TreePM code and Y. Maruyama for being
the first user of FDPS.  We are grateful to M. Tsubouchi for her help
in managing the FDPS development team. This research used
computational resources of the K computer provided by the RIKEN
Advanced Institute for Computational Science through the HPCI System
Research project (Project ID:ra000008). Part of the research covered
in this paper research was funded by MEXT's program for the
Development and Improvement for the Next Generation Ultra High-Speed
Computer System, under its Subsidies for Operating the Specific
Advanced Large Research Facilities. Numerical computations were in
part carried out on Cray XC30 at Center for Computational
Astrophysics, National Astronomical Observatory of Japan.


\begin{thebibliography}{}
\bibitem[Abrahama et al.(2014)]{2014GROMACS}
Abrahama,~M.~J., Murtolad,~T., Schulzb,~R., Palla,~S., Smithb,~J., Hessa,~B. \& Lindahl.,~E.\ 2015, SoftwareX, 1, 19
\bibitem[Asphaug \& Reufer(2014)]{2014NatGe...7..564A}
{{Asphaug},~E. \& {Reufer},~A.}\ 2014, Nature Geoscience, 7, 564
\bibitem[Bagla(2002)]{2002JApA...23..185B}
{Bagla},~J.~S.\ 2002, JApA, 23, 185
\bibitem[{Balsara}(1995)]{1995JCoPh.121..357B}
{Balsara},~D.~S.\ 1995, JCoPh, 121, 357
\bibitem[Barnes \& Hut(1986)]{1986Natur.324..446B}
{Barnes},~J. and {Hut},~P.\ 1986, Nature, 324, 446
\bibitem[Barnes(1990)]{1990JCoPh..87..161B}
{Barnes},~J.\ 1990, JCoPh, 87, 161
\bibitem[{B{\'e}dorf} et al.(2012)]{2012JCoPh.231.2825B}
{B{\'e}dorf},~J., {Gaburov},~E. \& {Portegies Zwart},~S.\ 2012, JCoPh, 231, 2825
\bibitem[B{\'e}dorf et al.(2014)]{Bedorf:2014:PGT:2683593.2683600}
B{\'e}dorf,~J., Gaburov,~E., Fujii,~M., Nitadori,~K., Ishiyama,~T., \& Portegies Zwart,~S.\ 2014, Proceedings of the International Conference for High Performance Computing, Networking, Storage and Analysis, IEEE Press, 54
\bibitem[Benz et al.(1986)]{1986Icar...66..515B}
{{Benz},~W., {Slattery},~W.~L. \& {Cameron},~A.~G.~W.}\ 1986, \icarus, 66, 515
\bibitem[Blackston \& Suel(1997)]{Blackston:1997:HPE:509593.509597}
Blackston,~D., \& Suel,~T.\ 1997, Proc. of the ACM/IEEE Conf. on Supercomputing, ACM, 1
\bibitem[Bode et al.(2000)]{2000ApJS..128..561B}
{Bode},~P., {Ostriker},~J.~P. \& {Xu},~G.\ 2000, ApJS, 128, 561
\bibitem[Brooks et al. (2009)]{2009CHARMM}
Brooks,~B., et al.\ 2009, J. Comp. Chem. 30, 1545
\bibitem[Cameron \& Ward(1976)]{1976LPI.....7..120C}
{{Cameron},~A.~G.~W. \& {Ward},~W.~R.}\ 1976, Lunar and Planetary Science Conference, 7, 120
\bibitem[Canup et al.(2013)]{2013Icar..222..200C}
{{Canup},~R.~M., {Barr},~A.~C. \& {Crawford},~D.~A.}\ 2013, \icarus, 222, 200
\bibitem[Case et al.(2015)]{2015AMBER}
Case,~D.~A., et al. 2015, AMBER 2015, (San Francisco:University of California)
\bibitem[Dehnen \& Aly(2012)]{2012MNRAS.425.1068D}
{{Dehnen},~W. \& {Aly},~H.}\ 2012, \mnras, 425, 1068
\bibitem[Dehnen(2000)]{2000ApJ...536L..39D}
Dehnen,~W.\ 2000, \apjl, 536, L39
\bibitem[Dubinski(1996)]{1996NewA....1..133D}
{Dubinski},~J.\ 1996, NewA, 1, 133,
\bibitem[Dubinski et al.(2004)]{2004NewA....9..111D}
Dubinski,~J., Kim,~J., Park,~C. \& Humble,~R.\ 2004, NewA, 9, 111
\bibitem[Fujii et al.(2011)]{2011ApJ...730..109F}
{Fujii},~M.~S., {Baba},~J., {Saitoh},~T.~R., {Makino},~J., {Kokubo},~E., \& {Wada},~K.\ 2011, ApJ, 730, 109
\bibitem[Gaburov et al.(2009)]{2009NewA...14..630G}
{Gaburov},~E., {Harfst},~S. \& {Portegies Zwart},~S.\ 2009, NewA, 14, 630
\bibitem[Goodale et al.(2003)]{2003Cactus}
Goodale,~T., Allen,~G., Lanfermann,~G., Mass{\'o},~J., Radke,~T., Seidel,~E. \& Shalf,~J.\ 2003, 5th International Conference, Lecture Notes in Computer Science, 2565
\bibitem[Hamada et al.(2009a)]{Hamada:2009:THN:1654059.1654123}
Hamada,~T., Narumi,~T., Yokota,~R., Yasuoka,~K., Nitadori,~K., \& Taiji,~M.\ 2009, Proceedings of the Conference on High Performance Computing Networking, Storage and Analysis, 62, 1  
\bibitem[Hamada et al.(2009b)]{hamada2009novel}
Hamada,~T., Nitadori,~K., Benkrid,~K., Ohno,~Y., Morimoto,~G., Masada,~T., Shibata,~Y., Oguri,~K. \& Taiji,~M.\ 2009, Computer Science-Research and Development, 24, 21
\bibitem[Hamada \& Nitadori(2010)]{Hamada:2010:TAN:1884643.1884644}
Hamada,~T., \& Nitadori,~K.\ 2010, Proceedings of the ACM/IEEE International Conference for High Performance Computing, Networking, Storage and Analysis, IEEE Computer Society, 1
\bibitem[{Hartmann} \& {Davis}(1975)]{1975Icar...24..504H}
{{Hartmann},~W.~K. \& {Davis},~D.~R.}\ 1975, \icarus, 24, 504
\bibitem[{Hernquist}(1990)]{1990ApJ...356..359H}
{Hernquist},~L.\ 1990, ApJ, 356, 359
\bibitem[Hockney \& Eastwood(1988)]{hockney1988computer}
Hockney,~R.~W. \& Eastwood,~J.~W.\ 1988, Computer Simulation Using Particles, CRC Press
\bibitem[Ishiyama et al.(2009)]{2009PASJ...61.1319I}
{Ishiyama},~T., {Fukushige},~T. \& {Makino},~J.\ 2009, PASJ, 61, 1319
\bibitem[Ishiyama et al.(2012)]{Ishiyama:2012:PAN:2388996.2389003}
Ishiyama,~T., Nitadori,~K., \& Makino, J.\ 2012, Proceedings of the International Conference on High Performance Computing, Networking, Storage and Analysis, 5, 1
\bibitem[Iwasawa et al.(2015)]{2015FDPS}
Iwasawa,~M., Tanikawa,~A., Hosono,~N., Nitadori,~K., Muranushi,~T. \& Makino,~J.\ 2015, WOLFHPC '15, 1, 1
\bibitem[Navarro et al.(1996)]{1996ApJ...462..563N}
{{Navarro},~J.~F., {Frenk},~C.~S. \& {White},~S.~D.~M.}\ 1996, ApJ, 462, 563
\bibitem[Nitadori et al.(2006)]{2006NewA...12..169N}
{Nitadori},~K., {Makino},~J. \& {Hut},~P.\ 2006, NewA, 12, 169
\bibitem[{Makino}(1991)]{1991PASJ...43..859M}
{Makino},~J.\ 1991, PASJ, 43, 859
\bibitem[Makino et al.(2003)]{2003PASJ...55.1163M}
Makino,~J., Fukushige,~T., Koga,~M., \& Namura,~K.\ 2003, \pasj, 55, 1163
\bibitem[Makino(2004)]{2004PASJ...56..521M}
{Makino},~J.\ 2004, \pasj, 56, 521  
\bibitem[Monaghan(1992)]{1992ARA&A..30..543M}
{Monaghan},~J.~J.\ 1992, ARA\&A, 30, 543
\bibitem[{Monaghan}(1997)]{1997JCoPh.136..298M}
{Monaghan},~J.~J.\ 1997, J. Comp. Phys., 136, 298
\bibitem[Murotani et al.(2014)]{2014Murotani}
Murotani,~K., et al.\ 2014, Journal of Advanced Simulation in Science and Engineering, 1, 16
\bibitem[Phillips et al.(2005)]{2005NAMD}
Phillips,~J., et al.\ 2005 J. Comp. Chem., 26, 1781
\bibitem[Plimpton(1995)]{1995LAMMPS}
Plimpton,~S.\ 1995, J. Comp. Phys., 117, 1
\bibitem[{Rosswog}(2009)]{2009NewAR..53...78R}
{Rosswog},~S.\ 2009, NewAR, 53, 78
\bibitem[Salmon \& Warren(1994)]{1994JCoPh.111..136S}
{Salmon},~J.~K. \& {Warren},~M.~S.\ 1994, 111, 136,
\bibitem[Schuessler \& Schmitt(1981)]{1981A&A....97..373S}
{Schuessler},~I. \& {Schmitt}, D.\ 1981, A\&A, 97, 3735
\bibitem[Shaw et al.(2014)]{GB14}
Shaw,~D.~E., et al.\ 2014, Proceedings of the International Conference on High Performance Computing, Networking, Storage and Analysis, 41
\bibitem[Springel(2005)]{2005MNRAS.364.1105S}
{Springel},~V.\ 2005, \mnras, 364, 1105
\bibitem[Springel et al.(2005)]{2005Natur.435..629S}
{Springel},~V., et al., 2005, \nat, 435, 629
\bibitem[{Springel}(2010)]{2010ARA&A..48..391S}
{Springel},~V.\ 2010, ARA\&A, 48, 391
\bibitem[Sugimoto et al.(1990)]{1990Natur.345...33S}
Sugimoto,~D., et al.\ 1990, \nat, 345, 33
\bibitem[Tanikawa et al.(2012)]{2012NewA...17...82T}
{Tanikawa},~A., {Yoshikawa},~K., {Okamoto},~T. \& {Nitadori},~K.\ 2012, NewA, 17, 82
\bibitem[Tanikawa et al.(2013)]{2013NewA...19...74T}
{Tanikawa},~A., {Yoshikawa},~K., {Nitadori},~K. \& {Okamoto},~T.\ 2013, NewA, 19, 74
\bibitem[Teodoro et al.(2014)]{2014LPI....45.2703T}
{{Teodoro},~L.~F.~A., {Warren},~M.~S., {Fryer},~C., {Eke},~V. \& {Zahnle},,K.}\ 2014, Lunar and Planetary Science Conference, 45, 2703
\bibitem[{Wadsley} et al.(2004)]{2004NewA....9..137W}
{Wadsley},~J.~W., {Stadel},~J. \& {Quinn},~T.\ 2004, NewA, 9, 137
\bibitem[Warren \& Salmon(1995)]{1995CoPhC..87..266W}
{Warren},~M.~S., \& {Salmon},~J.~K.\ 1995, Computer Physics Communications, 87, 266
\bibitem{confscWarrenSBGSW97}
M.~S.~Warren, J.~K.~Salmon, D.~J.~Becker, M.~P.~Goda, T.~L.~Sterling, \& W.~Winckelmans,\ 1997, Proceedings of the International Conference on High Performance Computing, Networking, Storage and Analysis, 61
\bibitem[Widrow \& Dubinski(2005)]{2005ApJ...631..838W}
{{Widrow},~L.~M. \& {Dubinski},~J.}\ 2005, ApJ, 631, 838
\bibitem[Xu(1995)]{1995ApJS...98..355X}
{Xu},~G.\ 1995, ApJS, 98, 355
\bibitem[Yamada et al.(2015)]{2015LexADV_EMPS}
Yamada,~T., Mitsume,~N., Yoshimura,~S. \& Murotani,~K.\ 2015, COUPLED PROBLEMS 2015
\bibitem[Yoshikawa \& Fukushige(2005)]{2005PASJ...57..849Y}
{Yoshikawa},~K. \& {Fukushige},~T.\ 2005, \pasj, 57, 849
  
\end{thebibliography}
\end{document}